\input{aipcheck}

\documentclass[
  ,numberedheadings 
  ]
  {aipproc}

\layoutstyle{8x11single}

\begin{document}

\title{Search for a Realistic Orbifold Grand Unification\footnote{
Talk presented at International Workshop on Grand Unified Theories: Current Status and Future Prospects (GUT07), 
December 17-19 2007, Kusatsu, Japan}}


\author{Yoshiharu KAWAMURA}{
  address={Department of Physics, Shinshu University, Matsumoto 390-8621, Japan}
}



\begin{abstract}
We review the prototype model of a grand unified theory on the orbifold $S^1/Z_2$  
and discuss topics related to the choice of boundary conditions;
the dynamical rearrangement of gauge symmetry and the equivalence classes of BCs.
We explore a family unification scenario by orbifolding.
\end{abstract}

\maketitle

\section{Introduction}

The standard model (SM) has been established as an effective theory below the weak scale.
Most people, however, believe that the SM cannot be an ultimate theory of nature 
because it has several problems.
I list some of them here.
\begin{flushleft}
\begin{tabular}{|l|}
\hline
~~~ {\bf Problems in the SM}\\
1. Why is the electric charge quantized?\\
2. What is the origin of anomaly free sets for matter fields?\\
3. The SM contains many independent parameters.\\
4. Naturalness problem.\\
\hline
\end{tabular}
\end{flushleft}
The situation of the first three problems has been improved in grand unified theories (GUTs)\cite{GUT}.
The naturalness problem is technically and partially solved by the introduction of supersymmetry (SUSY)\cite{NHK}.
Hence the grand unification and the SUSY are very attractive concepts, and we would like to go with them.

Actually, the SUSY grand unification scenario has attracted much attention as the physics beyond the minimal
SUSY extension of the SM (MSSM), 
since it was shown that the gauge coupling unification occurs on the basis of the MSSM
with a big desert hypothesis between the TeV scale and the grand unification scale\cite{GCU}.
The SUSY GUT\cite{SUSYGUT} has become a candidate as the proper theory beyond the MSSM.
We, however, encounter several problems towards the construction of a realistic model. 
I list some of them here.
\begin{flushleft}
\begin{tabular}{|l|}
\hline
~~~ {\bf Problems in SUSY GUT}\\
1. Triplet-doublet splitting problem.\\
2. Why is the proton so stable?\\
3. What is the origin of fermion mass hierarchy and mixing?\\
4. What is the origin of family?\\ \hline
\end{tabular}
\end{flushleft}
The first problem is what the breaking mechanism of a grand unified symmetry is such that
the triplet-doublet Higg mass splitting is naturally realized without fine-tuning among parameters.
For the first three problems, many intriguing ideas have been proposed, in most case, 
on the basis of {\it the extension of Higgs sector}.
The origin of the family replication has also been a big riddle and there have been several interesting proposals. 
But we have not arrived at final answers or well-established ones for these questions, yet.
Hence we would like to reconsider them from another angle, 
that is, by {\it the extension of our space-time structure}.

Now it is time to tell you our standpoint and our goal.
Our standpoint is that we shall adopt the SUSY grand unification scenario, and
our goal is to construct a realistic GUT with an extra space.
But we still have a long way to go there, and so the goal in this article is 
to introduce the construction of a GUT on the orbifold $S^1/Z_2$ 
and to discuss topics related to the choice of boundary conditions (BCs) 
on $S^1/Z_2$ and to suggest an origin of family.
These studies will help us on the constuction of a realistic model.

The content of this article is as follows.
First I explain the orbifold breaking mechanism and the prototype model of an orbifold GUT.
We will find what excellent features this type of model has.
Next I discuss topics related to the BCs on the orbifold, i.e., the dynamical rearrangement of
gauge symmetry and the equivalence classes of BCs on $S^1/Z_2$.
Then we study a family unification scenario by $Z_2$ orbifolding.
Finally I will give you a brief summary.

\section{Orbifold Grand Unified Theory}

\subsection{Orbifold and orbifold breaking}

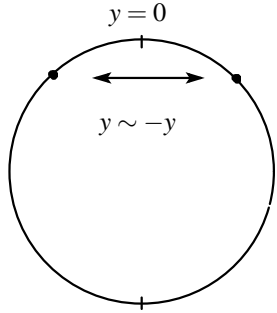
\begin{figure}[hbtp]
\caption{Orbifold $S^1/Z_2$}
\label{F1.S1Z2}
\unitlength 0.1in
\begin{picture}( 13.8200, 17.4000)(  4.5900,-17.8000)
%
\special{pn 13}%
\special{ar 1150 970 692 692  0.2763273 6.2831853}%
\special{ar 1150 970 692 692  0.0000000 0.2484874}%
%
\special{pn 13}%
\special{pa 1150 260}%
\special{pa 1150 320}%
\special{fp}%
%
\special{pn 13}%
\special{pa 1150 1630}%
\special{pa 1150 1690}%
\special{fp}%
%
\special{pn 20}%
\special{sh 1}%
\special{ar 1640 480 10 10 0  6.28318530717959E+0000}%
\special{sh 1}%
\special{ar 1650 480 10 10 0  6.28318530717959E+0000}%
\special{sh 1}%
\special{ar 1650 480 10 10 0  6.28318530717959E+0000}%
\special{sh 1}%
\special{ar 1650 490 10 10 0  6.28318530717959E+0000}%
%
\special{pn 20}%
\special{sh 1}%
\special{ar 680 460 10 10 0  6.28318530717959E+0000}%
\special{sh 1}%
\special{ar 690 460 10 10 0  6.28318530717959E+0000}%
\special{sh 1}%
\special{ar 690 460 10 10 0  6.28318530717959E+0000}%
\special{sh 1}%
\special{ar 690 470 10 10 0  6.28318530717959E+0000}%
\put(9.8000,-19.5000){\makebox(0,0)[lb]{$y=\pi R$}}%
\put(9.8000,-2.1000){\makebox(0,0)[lb]{$y=0$}}%
\put(9.3000,-7.8000){\makebox(0,0)[lb]{$y \sim -y$}}%
%
\special{pn 13}%
\special{pa 940 480}%
\special{pa 1460 480}%
\special{fp}%
\special{sh 1}%
\special{pa 1460 480}%
\special{pa 1394 460}%
\special{pa 1408 480}%
\special{pa 1394 500}%
\special{pa 1460 480}%
\special{fp}%
%
\special{pn 13}%
\special{pa 940 480}%
\special{pa 910 480}%
\special{fp}%
\special{sh 1}%
\special{pa 910 480}%
\special{pa 978 500}%
\special{pa 964 480}%
\special{pa 978 460}%
\special{pa 910 480}%
\special{fp}%
\end{picture}%
\end{figure}

{\it An orbifold is a space obtained by dividing a manifold with
some discrete transformation group, and the space has fixed points.}
Here, fixed points are points that transform into themselves under the discrete transformation.\footnote{
Orbifolds were initially utilized on the construction of
4-dimensional heterotic string models\cite{Orbifold,C&K}.}
The simple example of an orbifold is $S^1/Z_2$ and it is obtained by dividing a circle $S^1$
(whose radius is $R$) with the $Z_2$ transformation $y \to -y$, which is shown in Figure \ref{F1.S1Z2}.
As the point $y$ is identified with the point $-y$ on $S^1/Z_2$,
the space is regarded as a line segment (interval) whose length is $\pi R$.
The both end points $y = 0$ and $\pi R$ are fixed points under the $Z_2$ transformation.
For operations defined by
\begin{eqnarray}
Z_2: y \to -y ,~~ Z'_2: y \to 2 \pi R - y ,~~ T: y \to y + 2\pi R,
\label{Z2S}
\end{eqnarray}
the following relations hold:
\begin{eqnarray}
Z_2^2 = {Z'_2}^2 = I ,~~ T= Z_2 Z'_2  ,
\label{Z2S-rel}
\end{eqnarray}
where $I$ is the identity operation.
The operation $Z'_2$ is the reflection at the end point $y = \pi R$.

Let us adopt {\it the brane world scenario} with the help of orbifold fixed points.
We assume that the space-time is factorized into
a product of 4-dimensional Minkowski space and the orbifold $S^1/Z_2$, which is shown in Figure \ref{F2.Bworld}.
Those coordinates are denoted by $x$ and $y$, respectively. 
Our 4-dimensional world is assumed to be sitting on one of the fixed points. 
\begin{figure}[hbtp]
\caption{Brane world}
\label{F2.Bworld}
\unitlength 0.1in
\begin{picture}( 19.4000, 23.4000)(  7.7000,-30.2000)
%
\special{pn 13}%
\special{pa 920 1310}%
\special{pa 1550 680}%
\special{pa 1550 2230}%
\special{pa 930 2850}%
\special{pa 920 2830}%
\special{pa 920 1310}%
\special{fp}%
%
\special{pn 13}%
\special{pa 2080 1310}%
\special{pa 2710 680}%
\special{pa 2710 2230}%
\special{pa 2090 2850}%
\special{pa 2080 2830}%
\special{pa 2080 1310}%
\special{fp}%
%
\special{pn 13}%
\special{pa 920 2860}%
\special{pa 2080 2850}%
\special{fp}%
\put(7.7000,-31.0000){\makebox(0,0)[lb]{$y=0$}}%
\put(19.4000,-31.0000){\makebox(0,0)[lb]{$y=\pi R$}}%
\put(13.4000,-31.9000){\makebox(0,0)[lb]{$S^1/Z_2$}}%
\put(10.2000,-15.2000){\makebox(0,0)[lb]{$M^4$}}%
\put(22.0000,-15.1000){\makebox(0,0)[lb]{$M^4$}}%
\end{picture}%
\end{figure}
There exist two kinds of fields in the 5-dimensional space-time.
One is the brane field which exists only at the 4-dimensional boundary 
and the other is the bulk field which can go to the fifth direction.
On the orbifold, the point $y$ is identified with the points
$-y$ and $2 \pi R - y$, but each bulk field do not necessarily take an identical value at these points.
Let the bulk field $\Phi(x,y)$ be a multiplet of some transformation group $G$ and
the Lagrangian density $\mathcal{L}$ be invariant under the transformation 
$\Phi(x, y) \to \Phi'(x, y)=T_{\Phi}\Phi(x,y)$ such that $\displaystyle{\mathcal{L}(\Phi(x, y)) = \mathcal{L}(\Phi'(x, y))}$
where $T_{\Phi}$ is a representation matrix of $G$.
If we require that the $\mathcal{L}$ should be single-valued on $M^4 \times (S^1/Z_2)$, i.e.,
\begin{eqnarray}
\mathcal{L}(\Phi(x, y)) = \mathcal{L}(\Phi(x, -y)) = \mathcal{L}(\Phi(x, 2 \pi R -y)),
\label{L-single}
\end{eqnarray}
the following BCs are allowed, 
\begin{eqnarray}
\Phi(x, -y) = T_{\Phi}[P_0] \Phi(x, y) , ~~
\Phi(x, 2 \pi R -y) = T_{\Phi}[P_1] \Phi(x, y) , ~~
\Phi(x, y + 2 \pi R) = T_{\Phi}[U] \Phi(x, y) ,
\label{BCs-Phi}
\end{eqnarray}
where $T_{\Phi}[P_0]$, $T_{\Phi}[P_1]$ and $T_{\Phi}[U]$ represent appropriate representation matrices of $G$
and they satisfy the relations:
\begin{eqnarray}
T_{\Phi}[P_0]^2 = T_{\Phi}[P_1]^2 = {\mathcal{I}} , ~~
T_{\Phi}[U] = T_{\Phi}[P_1] T_{\Phi}[P_0] .
\label{Phi-rel}
\end{eqnarray}
Here, the $\mathcal{I}$ stands for the unit matrix.
The matrices $P_0$, $P_1$ and $U$ are the representation
matrices (up to sign factors) of the fundamental representation of the $Z_2$, $Z'_2$ and $T$ transformations, respectively.
The eigenvalues of $T_{\Phi}[P_0]$ and $T_{\Phi}[P_1]$
are interpreted as $Z_2$ parities (denoted as $(\mathcal{P}_0, \mathcal{P}_1)$) 
of each component for the fifth coordinate flip.
Because the assignment of $Z_2$ parities determines BCs of each multiplet 
(or transformation properties under the $Z_2$ reflections) as given by (\ref{BCs-Phi}), 
we use $\lq\lq$$Z_2$ parities" or $\lq\lq$$Z_2$ parity assignment"
as a parallel expression of $\lq\lq$BCs on $S^1/Z_2$".

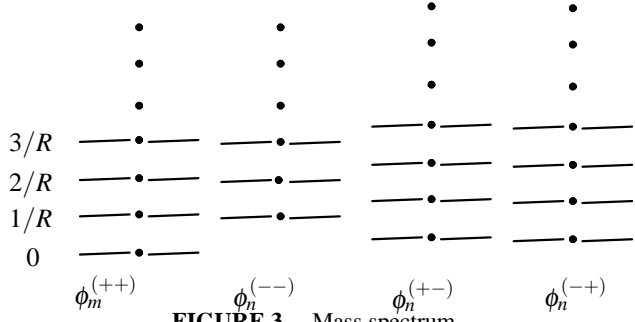
\begin{figure}[hbtp]
\caption{Mass spectrum}
\label{F3.KKmass}
\unitlength 0.1in
\begin{picture}( 32.6000, 14.4000)(  0.9000,-16.2000)
%
\special{pn 13}%
\special{pa 460 890}%
\special{pa 720 880}%
\special{fp}%
%
\special{pn 13}%
\special{pa 820 890}%
\special{pa 1080 880}%
\special{fp}%
%
\special{pn 20}%
\special{sh 1}%
\special{ar 770 880 10 10 0  6.28318530717959E+0000}%
\special{sh 1}%
\special{ar 770 880 10 10 0  6.28318530717959E+0000}%
%
\special{pn 13}%
\special{pa 460 1090}%
\special{pa 720 1080}%
\special{fp}%
%
\special{pn 13}%
\special{pa 820 1090}%
\special{pa 1080 1080}%
\special{fp}%
%
\special{pn 20}%
\special{sh 1}%
\special{ar 770 1080 10 10 0  6.28318530717959E+0000}%
\special{sh 1}%
\special{ar 770 1080 10 10 0  6.28318530717959E+0000}%
%
\special{pn 13}%
\special{pa 460 1280}%
\special{pa 720 1270}%
\special{fp}%
%
\special{pn 13}%
\special{pa 820 1280}%
\special{pa 1080 1270}%
\special{fp}%
%
\special{pn 20}%
\special{sh 1}%
\special{ar 770 1270 10 10 0  6.28318530717959E+0000}%
\special{sh 1}%
\special{ar 770 1270 10 10 0  6.28318530717959E+0000}%
%
\special{pn 13}%
\special{pa 460 1480}%
\special{pa 720 1470}%
\special{fp}%
%
\special{pn 13}%
\special{pa 820 1480}%
\special{pa 1080 1470}%
\special{fp}%
%
\special{pn 20}%
\special{sh 1}%
\special{ar 770 1470 10 10 0  6.28318530717959E+0000}%
\special{sh 1}%
\special{ar 770 1470 10 10 0  6.28318530717959E+0000}%
%
\special{pn 13}%
\special{pa 1990 810}%
\special{pa 2250 800}%
\special{fp}%
%
\special{pn 13}%
\special{pa 2350 810}%
\special{pa 2610 800}%
\special{fp}%
%
\special{pn 20}%
\special{sh 1}%
\special{ar 2300 800 10 10 0  6.28318530717959E+0000}%
\special{sh 1}%
\special{ar 2300 800 10 10 0  6.28318530717959E+0000}%
%
\special{pn 13}%
\special{pa 1990 1010}%
\special{pa 2250 1000}%
\special{fp}%
%
\special{pn 13}%
\special{pa 2350 1010}%
\special{pa 2610 1000}%
\special{fp}%
%
\special{pn 20}%
\special{sh 1}%
\special{ar 2300 1000 10 10 0  6.28318530717959E+0000}%
\special{sh 1}%
\special{ar 2300 1000 10 10 0  6.28318530717959E+0000}%
%
\special{pn 13}%
\special{pa 1990 1200}%
\special{pa 2250 1190}%
\special{fp}%
%
\special{pn 13}%
\special{pa 2350 1200}%
\special{pa 2610 1190}%
\special{fp}%
%
\special{pn 20}%
\special{sh 1}%
\special{ar 2300 1190 10 10 0  6.28318530717959E+0000}%
\special{sh 1}%
\special{ar 2300 1190 10 10 0  6.28318530717959E+0000}%
%
\special{pn 13}%
\special{pa 1990 1400}%
\special{pa 2250 1390}%
\special{fp}%
%
\special{pn 13}%
\special{pa 2350 1400}%
\special{pa 2610 1390}%
\special{fp}%
%
\special{pn 20}%
\special{sh 1}%
\special{ar 2300 1390 10 10 0  6.28318530717959E+0000}%
\special{sh 1}%
\special{ar 2300 1390 10 10 0  6.28318530717959E+0000}%
%
\special{pn 13}%
\special{pa 2730 820}%
\special{pa 2990 810}%
\special{fp}%
%
\special{pn 13}%
\special{pa 3090 820}%
\special{pa 3350 810}%
\special{fp}%
%
\special{pn 20}%
\special{sh 1}%
\special{ar 3040 810 10 10 0  6.28318530717959E+0000}%
\special{sh 1}%
\special{ar 3040 810 10 10 0  6.28318530717959E+0000}%
%
\special{pn 13}%
\special{pa 2730 1020}%
\special{pa 2990 1010}%
\special{fp}%
%
\special{pn 13}%
\special{pa 3090 1020}%
\special{pa 3350 1010}%
\special{fp}%
%
\special{pn 20}%
\special{sh 1}%
\special{ar 3040 1010 10 10 0  6.28318530717959E+0000}%
\special{sh 1}%
\special{ar 3040 1010 10 10 0  6.28318530717959E+0000}%
%
\special{pn 13}%
\special{pa 2730 1210}%
\special{pa 2990 1200}%
\special{fp}%
%
\special{pn 13}%
\special{pa 3090 1210}%
\special{pa 3350 1200}%
\special{fp}%
%
\special{pn 20}%
\special{sh 1}%
\special{ar 3040 1200 10 10 0  6.28318530717959E+0000}%
\special{sh 1}%
\special{ar 3040 1200 10 10 0  6.28318530717959E+0000}%
%
\special{pn 13}%
\special{pa 2730 1410}%
\special{pa 2990 1400}%
\special{fp}%
%
\special{pn 13}%
\special{pa 3090 1410}%
\special{pa 3350 1400}%
\special{fp}%
%
\special{pn 20}%
\special{sh 1}%
\special{ar 3040 1400 10 10 0  6.28318530717959E+0000}%
\special{sh 1}%
\special{ar 3040 1400 10 10 0  6.28318530717959E+0000}%
%
\special{pn 13}%
\special{pa 1200 900}%
\special{pa 1460 890}%
\special{fp}%
%
\special{pn 13}%
\special{pa 1560 900}%
\special{pa 1820 890}%
\special{fp}%
%
\special{pn 20}%
\special{sh 1}%
\special{ar 1510 890 10 10 0  6.28318530717959E+0000}%
\special{sh 1}%
\special{ar 1510 890 10 10 0  6.28318530717959E+0000}%
%
\special{pn 13}%
\special{pa 1200 1100}%
\special{pa 1460 1090}%
\special{fp}%
%
\special{pn 13}%
\special{pa 1560 1100}%
\special{pa 1820 1090}%
\special{fp}%
%
\special{pn 20}%
\special{sh 1}%
\special{ar 1500 1090 10 10 0  6.28318530717959E+0000}%
\special{sh 1}%
\special{ar 1500 1090 10 10 0  6.28318530717959E+0000}%
%
\special{pn 13}%
\special{pa 1200 1290}%
\special{pa 1460 1280}%
\special{fp}%
%
\special{pn 13}%
\special{pa 1560 1290}%
\special{pa 1820 1280}%
\special{fp}%
%
\special{pn 20}%
\special{sh 1}%
\special{ar 1510 1280 10 10 0  6.28318530717959E+0000}%
\special{sh 1}%
\special{ar 1510 1280 10 10 0  6.28318530717959E+0000}%
\put(4.3000,-17.7000){\makebox(0,0)[lb]{$\phi^{(++)}_m$}}%
\put(12.6000,-17.8000){\makebox(0,0)[lb]{$\phi^{(--)}_n$}}%
\put(20.9000,-17.9000){\makebox(0,0)[lb]{$\phi^{(+-)}_n$}}%
\put(28.9000,-17.8000){\makebox(0,0)[lb]{$\phi^{(-+)}_n$}}%
%
\special{pn 20}%
\special{sh 1}%
\special{ar 770 700 10 10 0  6.28318530717959E+0000}%
\special{sh 1}%
\special{ar 770 700 10 10 0  6.28318530717959E+0000}%
%
\special{pn 20}%
\special{sh 1}%
\special{ar 770 480 10 10 0  6.28318530717959E+0000}%
\special{sh 1}%
\special{ar 770 480 10 10 0  6.28318530717959E+0000}%
%
\special{pn 20}%
\special{sh 1}%
\special{ar 770 290 10 10 0  6.28318530717959E+0000}%
\special{sh 1}%
\special{ar 770 290 10 10 0  6.28318530717959E+0000}%
%
\special{pn 20}%
\special{sh 1}%
\special{ar 1510 700 10 10 0  6.28318530717959E+0000}%
\special{sh 1}%
\special{ar 1510 700 10 10 0  6.28318530717959E+0000}%
%
\special{pn 20}%
\special{sh 1}%
\special{ar 1510 480 10 10 0  6.28318530717959E+0000}%
\special{sh 1}%
\special{ar 1510 480 10 10 0  6.28318530717959E+0000}%
%
\special{pn 20}%
\special{sh 1}%
\special{ar 1510 290 10 10 0  6.28318530717959E+0000}%
\special{sh 1}%
\special{ar 1510 290 10 10 0  6.28318530717959E+0000}%
%
\special{pn 20}%
\special{sh 1}%
\special{ar 2300 590 10 10 0  6.28318530717959E+0000}%
\special{sh 1}%
\special{ar 2300 590 10 10 0  6.28318530717959E+0000}%
%
\special{pn 20}%
\special{sh 1}%
\special{ar 2300 370 10 10 0  6.28318530717959E+0000}%
\special{sh 1}%
\special{ar 2300 370 10 10 0  6.28318530717959E+0000}%
%
\special{pn 20}%
\special{sh 1}%
\special{ar 2300 180 10 10 0  6.28318530717959E+0000}%
\special{sh 1}%
\special{ar 2300 180 10 10 0  6.28318530717959E+0000}%
%
\special{pn 20}%
\special{sh 1}%
\special{ar 3040 600 10 10 0  6.28318530717959E+0000}%
\special{sh 1}%
\special{ar 3040 600 10 10 0  6.28318530717959E+0000}%
%
\special{pn 20}%
\special{sh 1}%
\special{ar 3040 380 10 10 0  6.28318530717959E+0000}%
\special{sh 1}%
\special{ar 3040 380 10 10 0  6.28318530717959E+0000}%
%
\special{pn 20}%
\special{sh 1}%
\special{ar 3040 190 10 10 0  6.28318530717959E+0000}%
\special{sh 1}%
\special{ar 3040 190 10 10 0  6.28318530717959E+0000}%
\put(1.8000,-15.4000){\makebox(0,0)[lb]{$0$}}%
\put(0.9000,-13.7000){\makebox(0,0)[lb]{$1/R$}}%
\put(0.9000,-11.8000){\makebox(0,0)[lb]{$2/R$}}%
\put(0.9000,-9.7000){\makebox(0,0)[lb]{$3/R$}}%
\end{picture}%
\end{figure}

Let $\phi^{(\mathcal{P}_0 \mathcal{P}_1)}(x, y)$ be a component 
with definite $Z_2$ parities $(\mathcal{P}_0, \mathcal{P}_1)$ in $\Phi(x,y)$. 
The $\phi^{(\mathcal{P}_0 \mathcal{P}_1)}(x, y)$ is expanded as
\begin{eqnarray}
&~& \phi^{(++)}(x, y) = \frac{1}{\sqrt{\pi R}} \phi_0^{(++)}(x) + \sqrt{\frac{2}{\pi R}} 
\sum_{n=1}^{\infty} \phi_n^{(++)}(x) \cos \frac{ny}{R} ,
\label{phi++}\\
&~& \phi^{(+-)}(x, y) = \sqrt{\frac{2}{\pi R}} \sum_{n=1}^{\infty} \phi_n^{(+-)}(x) \sin \frac{ny}{R} ,
\label{phi+-}\\
&~& \phi^{(-+)}(x, y) = \sqrt{\frac{2}{\pi R}} \sum_{n=1}^{\infty} \phi_n^{(-+)}(x) \cos \frac{\left(n-\frac{1}{2}\right)y}{R} ,
\label{phi-+}\\
&~& \phi^{(--)}(x, y) = \sqrt{\frac{2}{\pi R}} \sum_{n=1}^{\infty} \phi_n^{(--)}(x) \sin \frac{\left(n-\frac{1}{2}\right)y}{R} ,
\label{phi--}
\end{eqnarray}
where $\pm$ indicates the eigenvalues $\pm1$.
In (\ref{phi++})--(\ref{phi--}), the coefficients $\phi_m^{(++)}(x)$ ($m=0,1,\dots$)
are 4-dimensional fields which acquire the masses $m/R$ when the $Z_2$ parities are $(+1, +1)$,
$\phi_n^{(--)}(x)$ ($n=1,\dots$) acquire the masses $n/R$ $(n=1,2,\dots)$ when the $Z_2$ parities are $(-1, -1)$, 
and $\phi_n^{(\pm\mp)}(x)$ ($n=1,\dots$) acquire the masses $(n-\frac{1}{2})/R$ when the $Z_2$ parities 
are $(\pm1, \mp1)$ upon compactification.
The mass spectrum is shown in Figure \ref{F3.KKmass}.
The point is that {\it $T_{\Phi}[P_a]$s are not necessarily proportional to the unit matrix and
massless fields $(\phi_0^{(++)}(x))$ called zero modes appear only in the components with even $Z_2$ parities.}
Different $Z_2$ parity assignment is allowed for each component and then
symmetries of $\mathcal{L}$ can be broken by the difference between the BCs of each component.\footnote{
Scherk and Schwarz proposed the mechanism of SUSY breaking by the difference between the BCs of bosons and fermions\cite{S&S}.}
Here we summarize our statement.
\begin{flushleft}
\begin{tabular}{|l|}
\hline
{\it Unless all components of the non-singlet field have common $Z_2$ parities,
a symmetry reduction occurs upon}\\ 
{\it  compactification because zero modes are absent in fields with an odd parity.}\\ \hline
\end{tabular}
\end{flushleft}
This type of symmetry breaking mechanism is called the $\lq\lq$orbifold breaking mechanism".
The $Z_2$ orbifolding was used in superstring theory\cite{A} and heterotic $M$-theory\cite{H&W}.
In field theoretical models, it was applied to the reduction of global SUSY\cite{M&P} and then
to the reduction of gauge symmetry\cite{K1}.

\subsection{Orbifold Grand Unification}

I explain the prototype model of an orbifold SUSY GUT\cite{K2}.
The orbifold $S^1/Z_2 \times Z'_2$ was used in the original paper, 
but we use the $S^1/Z_2$ in this article.
Models with same particle contents are constructed using them, because
there is one-to-one correspondence between them.

Our 4-dimensional world is supposed to be the hypersurface fixed at $y=0$.
We assume that the 5-dimensional bulk fields consist of $SU(5)$ gauge supermultiplet $\mathcal{V}$ 
and two kinds of Higgs hypermultiplets $\mathcal{H}$ and $\overline{\mathcal{H}}$.
The components of $\mathcal{V}$ are given by
\begin{eqnarray}
\mathcal{V} = (A^{\alpha}_M(x,y), \lambda^{\alpha}_1(x,y), \lambda^{\alpha}_2(x,y), \sigma^{\alpha}(x,y)) ,
\label{5Dgauge-multiplet}
\end{eqnarray}
where the $A^{\alpha}_M(x,y)$ $(M = 0,1,2,3,5)$ is 5-dimensional gauge bosons, 
$\lambda^{\alpha}_1(x,y)$ and $\lambda^{\alpha}_2(x,y)$ are two kinds of gauginos and 
$\sigma^{\alpha}(x,y)$ is a real scalar field.
The $\mathcal{V}$ is decomposed, in 4 dimensions, to the vector superfield $V$ and the chiral superfield $\Sigma$ as
\begin{eqnarray}
V = (A^{\alpha}_{\mu}(x,y), \lambda^{\alpha}_1(x,y)) , ~~ 
\Sigma =(\sigma^{\alpha}(x,y) + iA^{\alpha}_y(x,y), \lambda^{\alpha}_2(x,y)) ,
\label{4Dgauge-multiplet}
\end{eqnarray}
where the index $\alpha$ indicates $SU(5)$ gauge generators $(T^{\alpha})$.
The components of $\mathcal{H}$ and $\overline{\mathcal{H}}$ are given by
\begin{eqnarray}
\mathcal{H} = (h(x,y), \tilde{h}(x,y); h^{c\dagger}(x,y), \tilde{h}^{c\dagger}(x,y)) , ~~
\overline{\mathcal{H}} = (\bar{h}(x,y), \tilde{\bar{h}}(x,y); \bar{h}^{c\dagger}(x,y), \tilde{\bar{h}}^{c\dagger}(x,y)) ,
\label{5DHiggs}
\end{eqnarray}
where the fields with tildes represent superpartners of Higgs bosons called $\lq$Higgsinos'.
The $\mathcal{H}$ and $\overline{\mathcal{H}}$ are decomposed into four kinds of chiral superfields
\begin{eqnarray}
H = (h(x,y), \tilde{h}(x,y)) , ~~ H^c = (h^{c}(x,y), \tilde{h}^{c}(x,y)) , ~~
\overline{H} = (\bar{h}(x,y), \tilde{\bar{h}}(x,y)) , ~~ \overline{H^c} = (\bar{h}^{c}(x,y), \tilde{\bar{h}}^{c}(x,y)) ,
\label{4DHiggs}
\end{eqnarray}
where $H$ and $\overline{H^c}$ transform as $\bf{5}$ representation
and $H^c$ and $\overline{H}$ transform as $\bar{\bf{5}}$ representation.

The BCs of each field are determined
up to an arbitrary sign factor called $\lq$intrinsic $Z_2$ parity',
if the representation matrices $P_0$ and $P_1$ are given.
The gauge bosons yield the following BCs,
\begin{eqnarray}
&~& A_{\mu}^{\alpha}(x, -y)T^{\alpha} = P_0 A_{\mu}^{\alpha}(x, y)T^{\alpha} P_0^{\dagger} ,~~ 
A_{y}^{\alpha}(x, -y)T^{\alpha} = - P_0 A_{y}^{\alpha}(x, y)T^{\alpha} P_0^{\dagger} ,
\label{gaugeBC1}\\
&~& A_{\mu}^{\alpha}(x, 2\pi R-y)T^{\alpha} = P_1 A_{\mu}^{\alpha}(x, y)T^{\alpha} P_1^{\dagger} ,~~ 
A_{y}^{\alpha}(x, 2\pi R-y)T^{\alpha} = - P_1 A_{y}^{\alpha}(x, y)T^{\alpha} P_1^{\dagger} .
\label{gaugeBC2}
\end{eqnarray} 
These BCs are consistent with the gauge covariance of the covariant derivative $D_M \equiv \partial_M - ig A_M(x,y)$.
When we choose the representation matrix: 
\begin{eqnarray}
P_0 = \mbox{diag}(+1, +1, +1, +1, +1) ,~~ P_1 = \mbox{diag}(-1, -1, -1, +1, +1) ,
\label{GUT-parity}
\end{eqnarray}
the $Z_2$ parities for the members of $\mathcal{V}$ are fixed as
\begin{eqnarray}
A_{\mu}^{a(++)} ,~~ A_{\mu}^{\hat{a}(+-)} ,~~  A_{y}^{a(--)} ,~~ A_{y}^{\hat{a}(-+)} , ~~
\lambda_1^{a(++)} ,~~ \lambda_1^{\hat{a}(+-)} ,~~ \lambda_2^{a(--)} ,~~ \lambda_2^{\hat{a}(-+)} ,~~ 
\sigma^{a(--)} ,~~ \sigma^{\hat{a}(-+)} ,
\label{gauge-parity}
\end{eqnarray} 
where the index $a$ indicates the gauge generators of the SM gauge group 
$G_{\rm SM} \equiv SU(3)_C \times SU(2)_L \times U(1)_Y$ and 
the index $\hat{a}$ indicates other generators.
We find that the MSSM gauge multiplet $V_{\rm SM} = (A_{\mu}^{a(++)}, \lambda_1^{a(++)})$ 
has even $Z_2$ parities and their zero modes (denoted as $A_{\mu}^a(x)$ and $\lambda^a(x)$) survive in our 4-dimensional world.
\begin{center}
\begin{tabular}{|l|}
\hline
$\displaystyle{\mathcal{V}(x, y) \stackrel{S^1/Z_2}\longrightarrow (A_{\mu}^a(x) , \lambda^a(x))}$
\\ \hline
\end{tabular}
\end{center}

The Higgs chiral superfields yield the following BCs,
\begin{eqnarray}
&~& H(x, -y) = \eta_{0H} P_0 H(x,y) , ~~ H^c(x, -y) = - \eta_{0H} P_0 H^c(x,y) ,
\nonumber \\
&~& \overline{H}(x, -y) = \eta_{0\bar{H}} P_0 \overline{H}(x,y) , ~~ 
\overline{H^c}(x, -y) = - \eta_{0\bar{H}} P_0 \overline{H^c}(x,y) ,
\label{HiggsBC1}\\
&~& H(x, 2\pi R-y) = \eta_{1H} P_1 H(x,y) , ~~ H^c(x, 2\pi R-y) = - \eta_{1H} P_1 H^c(x,y) ,
\nonumber \\
&~& \overline{H}(x, 2\pi R-y) = \eta_{1\bar{H}} P_1 \overline{H}(x,y) , ~~ 
\overline{H}^c(x, 2\pi R-y) = - \eta_{1\bar{H}} P_1 \overline{H^c}(x,y) ,
\label{HiggsBC2}
\end{eqnarray} 
where $\eta_{0H}$, $\eta_{1H}$, $\eta_{0\bar{H}}$ and $\eta_{1\bar{H}}$ are the intrinsic $Z_2$ parities.
They are consistent with the requirement that the kinetic term of Higgsinos should be
invariant under the $Z_2$ parity transformation.
When we choose (\ref{GUT-parity}) and $\eta_{0H} = \eta_{1H} = \eta_{0\bar{H}} = \eta_{1\bar{H}} = +1$,
the Higgs chiral superfields have the following $Z_2$ parities,
\begin{eqnarray}
H_C^{(+-)} , ~~ H_W^{(++)} , ~~ H_C^{c(-+)} , ~~ H_W^{c(--)} , ~~
\overline{H}_C^{(+-)} , ~~ \overline{H}_W^{(++)} , ~~ \overline{H}_C^{c(-+)} , ~~ \overline{H}_W^{c(--)} ,
\label{gauge-parity}
\end{eqnarray} 
where the $H_C$ and $H_W$ are colored triplet and weak doublet components, respectively.
We find that two kinds of weak Higgs chiral superfields $(H_W^{(++)}, \overline{H}_W^{(++)})$ have even $Z_2$ parities 
and their zero modes (denoted as $h_W(x)$, $\tilde{h}_W(x)$, $\bar{h}_W(x)$ and $\tilde{\bar{h}}_W(x)$) 
survive in our 4-dimensional world.
\begin{center}
\begin{tabular}{|l|}
\hline
$\displaystyle{(\mathcal{H}(x, y), \overline{\mathcal{H}}(x, y)) \stackrel{S^1/Z_2}\longrightarrow 
(h_W(x), \tilde{h}_W(x), \bar{h}_W(x), \tilde{\bar{h}}_W(x)) }$
\\ \hline
\end{tabular}
\end{center}

The point is that {\it because the colored components have an odd parity and no zero modes,
the gauge symmetry reduction such as $SU(5) \to G_{\mbox{\tiny{SM}}}$ occurs and 
the triplet-doublet mass splitting is elegantly realized through
the orbifold breaking}\footnote{
In 4-dimensional heterotic string models,
extra color triplets are projected out by the Wilson line mechanism\cite{IKNQ}.
Our symmetry reduction is regarded as the field-theoretical version.} 
{\it under the $Z_2$ parity assignment}:
$$\displaystyle{P_0 = \mbox{diag}(+1, +1, +1, +1, +1) ,~~ P_1 = \mbox{diag}(-1, -1, -1, +1, +1)} .$$

Our simple model was extended and studied intensively\cite{A&F,Hall&Nomura}.
Our brane fixed at $y=0$ has the $SU(5)$ gauge symmetry.
For the location of matter fields, there are many possibilities.
Some matter chiral superfields exist on our 4-dimensional brane as $SU(5)$ multiplets and others are in the bulk
as a member of hypermultiplets.
Three families can originate from both our brane fields and zero modes of bulk fields.
The Kaluza-Klein (KK) modes do not appear in our low-energy world,
because they have heavy masses of $O(1/R)$, 
which is the magnitude of the grand unification scale.

Let us come back to the problems in SUSY GUT.
The gauge symmetry reduction and triplet-doublet mass splitting have been elegantly 
realized through the orbifold breaking.
We need to consider the $\mu$ problem $\lq\lq$What is the origin of $\mu$ term in the MSSM?"
and the problem on the stability of proton.
These subjects were investigated from the structure of the theory\cite{Hall&Nomura}.
The dangerous $\mu$ term due to a brane interaction is forbidden by $U(1)_R$ symmetry.
Here, the $U(1)_R$ is the diagonal subgroup of two $U(1)$s from $SU(2)_R \times SU(2)_H$
where $SU(2)_R$ is the $R$ symmetry and $SU(2)_H$ is the flavor symmetry rotating the two kinds of Higgs hypermultiplets.
There are several scenarios to generate $\mu$ term with a suitable magnitude.
The $U(1)_R$ also forbids the dangerous proton decay processes via the KK modes of colored Higgsinos exchange.
The processes via the KK modes of X and Y gauge bosons exchange are suppressed 
if the matter fields with ${\bf 10}$ representation in the first two generations are the bulk fields.
The fermion mass hierarchy can be generated by the difference among the locations of matter fields, 
because the magnitude of each interaction among bulk fields 
and brane fields depends on the volume suppression factor\cite{N,Y}.

The origin of family will be discussed in section 4.
We return to the first problem.
We have found that the successful realization of mass splitting originates from the non-trivial
assignment of $Z_2$ parities or the suitable choice of BCs.
Hence, exactly speaking, the problem has not been completely solved and the following question remains.
\begin{flushleft}
\begin{tabular}{|l|}
\hline
~~~ {\bf Arbitrariness problem}\\
What is the origin of specific $Z_2$ parity assignment? 
Or what is the principle to determine BCs?\\ \hline
\end{tabular}
\end{flushleft}

\section{Topics on Boundary Conditions}

\subsection{Dynamical Rearrangement of Gauge Symmetry}

Bearing the arbitrariness problem in mind, 
we study the gauge transformation properties of BCs and the physical implications on $S^1/Z_2$\cite{HHHK}.
There are many kinds of $5 \times 5$ matrices that satisfy the relations (\ref{Phi-rel}). 
Diagonal matrices are written by
\begin{eqnarray}
&~& P_0 = \mbox{diag}(+1, +1, +1, +1, +1) ,~~ P_1 = \mbox{diag}(+1, +1, +1, +1, +1) ,
\nonumber\\
&~& P_0 = \mbox{diag}(+1, +1, +1, +1, +1) ,~~ P_1 = \mbox{diag}(-1, +1, +1, +1, +1) ,
\nonumber\\
&~& P_0 = \mbox{diag}(+1, +1, +1, +1, +1) ,~~ P_1 = \mbox{diag}(-1, -1, +1, +1, +1) ,
\nonumber\\
&~& P_0 = \mbox{diag}(+1, +1, +1, +1, +1) ,~~ P_1 = \mbox{diag}(-1, -1, -1, +1, +1) ,
\nonumber\\
&~& ~~~~~~~~\cdots 
\nonumber\\
&~& P_0 = \mbox{diag}(-1, -1, -1, -1, -1) ,~~ P_1 = \mbox{diag}(-1, -1, -1, -1, -1) 
\label{GUT-parities}
\end{eqnarray}
and one example of non-diagonal one is given by
\begin{eqnarray}
P_0 = \mbox{diag}(+1, +1, +1, +1, +1) , ~~ 
P_1 = 
\left(
\begin{array}{ccccc}
-\cos\pi p & 0 & 0 & -i\sin\pi p & 0 \\
 0 & -\cos\pi q & 0 & 0 & -i\sin\pi q \\
 0 & 0 & -1 & 0 & 0 \\
i\sin\pi p & 0 & 0 & \cos\pi p & 0 \\
 0 & i\sin\pi q & 0 & 0 & \cos\pi q 
\end{array}
\right) ,
\label{GUT-parities-off}
\end{eqnarray}
where $p$ and $q$ are arbitrary real numbers.
In this subsection, we show that some of them are related to by gauge transformations
and have the same physics content.

First we explain how a specific gauge transformation connects to different representation matrices.
Under the gauge transformation $\Phi(x,y) \to \Phi'(x,y) = T_{\Phi}[\Omega] \Phi(x,y)$,
the BCs change as
\begin{eqnarray}
\Phi'(x, -y) = T_{\Phi}[P'_0] \Phi'(x, y) , ~~
\Phi'(x, 2 \pi R -y) = T_{\Phi}[P'_1] \Phi'(x, y) , ~~
\Phi'(x, y + 2 \pi R) = T_{\Phi}[U'] \Phi'(x, y) ,
\label{BC's-vphi}
\end{eqnarray}
where $\Omega = \Omega(x,y)$ is a gauge transformation function and operators with primes are given by,
\begin{eqnarray}
P'_0 = \Omega(x,-y) \, P_0 \, \Omega^\dagger (x,y) , ~~
P'_1 = \Omega(x, 2\pi R -y) \, P_1 \, \Omega^\dagger (x,y)  , ~~
U' = \Omega(x,y+2\pi R) \,  U \, \Omega^\dagger (x,y) .
\label{BC3}
\end{eqnarray}
The point is that {\it the BCs do not necessarily agree with the original ones, i.e.,
$(P_0, P_1, U) \ne (P'_0, P'_1, U')$, for a singular gauge transformation.}
For example, for the $SU(2)$ gauge group, $(P_0, P_1) =(\tau_3, \tau_3)$ 
are transformed into $(P'_0, P'_1) =(\tau_3, - \tau_3)$
with the singular gauge transformation function $\displaystyle{\Omega = \exp\left(-i\frac{\tau_2}{2R}y\right)}$. 
Here, $\tau_i$s are Pauli matrices.
These two types of BCs are connected to by the gauge transformation and hence they should be equivalent, i.e.,
\begin{eqnarray}
(P_0 = \tau_3, P_1 = \tau_3, U=I) \sim (P'_0= \tau_3, P'_1 = - \tau_3, U' = -I) .
\label{equ}
\end{eqnarray}
If we use the equivalence relation (\ref{equ}), the following relations are derived for the $SU(5)$ group,
\begin{eqnarray}
\hspace{-1cm}&~& \left(P_0 = \mbox{diag}(+1, +1, -1, -1, -1) ,~ P_1 = \mbox{diag}(+1, +1, -1, -1, -1)\right) ,
\nonumber\\
\hspace{-1cm}&~& \sim \left(P_0 = \mbox{diag}(+1, +1, -1, -1, -1) ,~ P_1 = \mbox{diag}(+1, -1, +1, -1, -1)\right) ,
\nonumber\\
\hspace{-1cm}&~& \sim \left(P_0 = \mbox{diag}(+1, +1, -1, -1, -1) ,~ P_1 = \mbox{diag}(-1, -1, +1, +1, -1)\right) .
\label{GUT-equ}
\end{eqnarray}

Now let us check whether the equivalence holds or not from the viewpoint of physics.
On the mere face of it, there seem to exist two different statements.
We refer to the symmetry on the Fourier expansions of a mutiplet as $\lq\lq$the symmetry of BCs".
Components have different mode expansions if their BCs are different, and then we have the following statement.
\begin{flushleft}
\begin{tabular}{|l|}
\hline
S1. The symmetry of BCs, in general, differs from each other if mode expansions are different. \\ \hline
\end{tabular}
\end{flushleft}
We show it using a simple example.
We consider two theories that contain the
same particle content (a scalar field $h(x, y)$ whose representation is $\bf{5}$ of $SU(5)$) 
but the different BCs related to by the gauge transformation.
\begin{eqnarray}
&~& ({\mbox{BC}}1) ~~~ P_0 = \mbox{diag}(+1, +1, -1, -1, -1) ,~~ P_1 = \mbox{diag}(+1, +1, -1, -1, -1) ,
\label{BC1}\\
&~& ({\mbox{BC}}2) ~~~ P_0 = \mbox{diag}(+1, +1, -1, -1, -1) ,~~ P_1 = \mbox{diag}(+1, -1, +1, -1, -1) .
\label{BC2}
\end{eqnarray}
The mass terms upon compactification clearly differ from each other such that 
\begin{eqnarray}
&~& ({\mbox{BC}}1) ~~~ \int dy |\partial_y h(x,y)|^2 \Rightarrow 
\sum_{n=0}^{\infty}\left(\frac{n}{R}\right)^2 \left(|h_n^{1(++)}|^2  + |h_n^{2(++)}|^2\right)
+ \sum_{n=1}^{\infty}\left(\frac{n}{R}\right)^2 \left(|h_n^{3(--)}|^2  
+ |h_n^{4(--)}|^2 + |h_n^{5(--)}|^2\right) ,
\nonumber \\
&~& ~~~~~~~~~~~~~~~ \mbox{Symmetry of BCs}: SU(2) \times SU(3) \times U(1) ,
\label{Mass1}\\
&~& ({\mbox{BC}}2) ~~~ \int dy |\partial_y h(x,y)|^2 \Rightarrow 
\sum_{n=0}^{\infty}\left(\frac{n}{R}\right)^2 |h_n^{1(++)}|^2 
+ \sum_{n=1}^{\infty}\left(\frac{n}{2R}\right)^2 \left(|h_n^{2(+-)}|^2  + |h_n^{3(-+)}|^2\right)
\nonumber \\
&~& ~~~~~~~~~~~~~~~~~~~~~~~~~~~~~~~~~~~~~~~~~~~~~~~~ + \sum_{n=1}^{\infty}\left(\frac{n}{R}\right)^2 \left(|h_n^{4(--)}|^2 
+ |h_n^{5(--)}|^2\right) ,
\nonumber \\
&~& ~~~~~~~~~~~~~~~ \mbox{Symmetry of BCs}: SU(2) \times U(1) \times U(1) \times U(1) .
\label{Mass2}
\end{eqnarray}
Here, the symmetry of BCs is spanned by the generators which commutes with $(P_0, P_1, U)$
and differ from each other.
On the other hand, the following statement comes from the gauge principle, i.e.,
{\it the physics should be invariant under the gauge transformation.}
\begin{flushleft}
\begin{tabular}{|l|}
\hline
S2. The theories are equivalent and describe the same physics, if they are related to by gauge transformations.\\ \hline
\end{tabular}
\end{flushleft}
How can we bridge the gap between the different mode expansions and the gauge equivalence?
What is the physical symmetry?
The Hosotani mechanism answers the questions.
Let us consider gauge theories defined on a multiply connected space.
The Hosotani mechanism consists of the following several parts\cite{Hosotani}. 
\begin{flushleft}
\begin{tabular}{|l|}
\hline
~~~ {\bf Hosotani mechanism}\\
(i)  Wilson line phases are phases of $WU$ defined by
$\displaystyle{W U \equiv P \exp (ig  \int_C dy A_y) U}$.
Here,
$C$ is a non-contractible\\ 
~~~~ loop. The eigenvalues of $WU$ are gauge invariant and become
physical degrees of freedom. Wilson line phases\\ 
~~~~ cannot be gauged away and parametrize degenerate vacua at the classical level.\\
(ii) The degeneracy, in general, is lifted by quantum effects.
The physical vacuum is given by the configuration of\\
~~~~ Wilson line phases which minimizes the effective potential $V_{\rm eff}$.\\
(iii) If the configuration of the Wilson line phases is non-trivial,
the gauge symmetry is spontaneously broken or\\
~~~~~ restored by radiative corrections.
Nonvanishing vacuum expectation values (VEVs) of the Wilson line phases\\
~~~~~ give masses to those gauge fields in lower dimensions whose gauge symmetry is broken.
Some of matter\\ 
~~~~~ fields also acquire masses.\\
(iv)  Extra-dimensional components of gauge fields also become massive 
with the nontrivial $V_{\rm eff}$.\\
(v) Two sets of BCs for fields can be related to each other by a BCs-changing 
gauge transformation.
They are\\ 
~~~~ physically equivalent, even if the two sets have distinct symmetry of BCs.  
This defines equivalence classes\\
~~~~ of the BCs.   
The $V_{\rm eff}$ depends on the BCs so that the VEVs of the Wilson line phases
depend on the BCs.\\ 
~~~~ Physical symmetry of the theory is determined by the combination of the BCs and the VEVs
of the Wilson\\
~~~~ line phases. Theories in the same equivalence class of the BCs have
the same physical symmetry and physics\\
~~~~ content.\\
(vi) The physical symmetry of the theory is mostly dictated by the
matter content of the theory.\\
(vii)  The mechanism provides unification of gauge fields and Higgs scalar fields 
in the adjoint representation,\\ 
~~~~~~ namely the gauge-Higgs unification.\\ \hline
\end{tabular}
\end{flushleft}

The part (v) is just the answer of the first question. 
The equivalence between theories is understood from the gauge invariance of $\mathcal{L}$ as
\begin{eqnarray}
\mathcal{L}(\Phi(x,y))|_{(\langle A_y \rangle, P_0, P_1)} = \mathcal{L}(\Phi'(x,y))|_{(\langle A'_y \rangle, P'_0, P'_1)} 
\label{L-gaugeinv}
\end{eqnarray}
by using the gauge invariance of $V_{\rm eff}$, i.e., 
$V_{\rm eff}(A^{\rm bg}_M, P_0, P_1) = V_{\rm eff}(A'^{\rm bg}_M, P'_0, P'_1)$
in the background field gauge.
Here, $A^{\rm bg}_M$ is the background configuration of $A_M$.
For the above-mensioned example,
the $h(x,y)$ couples to $A_y$ in the gauge invariant fashion $|(\partial_y - ig A_y(x,y))h(x,y)|^2$ 
and the physical mass spectrum is determined by the combination of the BCs and the VEV of $A_y$.
To obtain the VEV of $A_y$, we need to find the minimum of $V_{\rm eff}$ for each theory.
After we find the minimum of $V_{\rm eff}$ and incorporate the VEV of $A_y$, 
we arrive at a same mass spectrum for two theories.
We refer to this phenomenon as the $\lq\lq$dynamical rearrangement of gauge symmetry".

We answer the second question in gauge theory defined on $M^4 \times (S^1/Z_2)$.
Dynamical Wilson line phases are given by $\{\theta^b = 2 \pi R g A_y^b ~,~ T^b \in \mathcal{H} \}$
where $T^b$s are generators which anticommute with $(P_0, P_1)$,
\begin{eqnarray}
\mathcal{H} = \{ T^b ~;~ \{T^b, P_0\} = \{T^b, P_1\} =0 ~ \} .
\label{decomposition2}
\end{eqnarray}
They correspond to the parts with even $Z_2$ parities.
Suppose that $V_{\rm eff}$ is minimized at $\langle  A_y \rangle$ 
such that $W \equiv \exp(ig 2\pi R  \langle A_y \rangle) \neq I$ with $(P_0, P_1, U)$.
Perform a BCs-changing gauge transformation
$\Omega = \exp\{i g (y + \alpha) \langle A_y \rangle\}$,
which brings $\langle A_y \rangle$ to $\langle {A'}_y \rangle = 0$.
Here, the gauge transformation of $A_y$ is given by 
$\displaystyle{A'_y = \Omega A_y \Omega^{\dagger} - \frac{i}{g} \Omega \partial_y \Omega^{\dagger}}$.
Under the transformation, the BCs change to
\begin{eqnarray}
(P_0^{\rm sym}, P_1^{\rm sym}, U^{\rm sym})
\equiv (P'_0, P'_1, U') = (e^{2ig\alpha\langle A_y \rangle} P_0,
e^{2ig(\alpha + \pi R)\langle A_y \rangle} P_1, e^{ig 2\pi R \langle A_y \rangle} U (=WU)) .
\label{equiv3}
\end{eqnarray}
Since the VEVs of $A_y'$ vanish in the new
gauge, the physical symmetry is spanned by the generators
which commute with $(P_0^{\rm sym}, P_1^{\rm sym}, U^{\rm sym})$;
\begin{eqnarray}
\mathcal{H}^{\rm sym} =  \{~T^a ~;~
[T^a, P_0^{\rm sym}] = [T^a, P_1^{\rm sym}] =0 ~ \} .
\label{decomposition3}
\end{eqnarray}
{\it The group, $H^{\rm sym}$, generated by $\mathcal{H}^{\rm sym}$ is the unbroken
physical symmetry of the theory.}
For our model discussed in the previous section, 
there is no radiative correction due to $A_y$, 
because all zero modes of $A_y$ are projected out by the BCs (\ref{GUT-parity}).
Hence the physical symmetry is same as the symmetry of BCs and the triplet-doublet splitting survives!

\subsection{Equivalence Classes of Boundary Conditions}

In formulating the theory on an orbifold, there are many possibilities for BCs.
We have found that some of them are gauge equivalent and arrive at the concept of
equivalence classes of BCs.
Now the arbitrariness problem is restated as:
$\lq\lq$what is the principle to select a specific or realistic equivalence class?"
It is tough to answer the question.
To provide information that is useful to solve the problem, we carry out the classification of equivalence
classes and evaluate the vacuum energy density\cite{HHK}.

For the classification of equivalence classes of BCs on $S^1/Z_2$,
we can show that each equivalence class has a diagonal representative for 
$T_{\Phi}[P_0]$ and $T_{\Phi}[P_1]$ for the $SU(N)$ gauge group.
The diagonal representatives are specified by three non-negative integers $(p, q, r)$ such that
\begin{eqnarray}
&~& \mbox{diag} P_0 = (\overbrace{+1, \cdots, +1, +1, \cdots, +1 ,
        -1, \cdots, -1, -1, \cdots, -1}^N) ~, \nonumber \\
&~& \mbox{diag} P_1 = (\underbrace{+1, \cdots, +1}_{p}, \underbrace{-1, \cdots, -1}_{q} ,
\underbrace{+1, \cdots, +1}_r, \underbrace{-1, \cdots, -1}_{s = N-p-q-r}) ~,
\label{pqr}
\end{eqnarray}
where $N \geq p, q, r, s \geq 0$.
We denote each theory with the BCs specified by $(p, q, r)$ as $[p; q, r; s]$.
The $P_0$ is interchanged with $P_1$ by the interchange between $q$ and $r$ such that
\begin{eqnarray}
[p; q, r; s] \leftrightarrow [p; r, q; s] .
\label{interchange}
\end{eqnarray}
Using the equivalence (\ref{equ}), we can derive the following equivalence relations in $SU(N)$ gauge theory;
\begin{eqnarray}
[p; q, r; s] &\sim& [p-1; q+1 , r+1; s-1] , ~~ \mbox{for} ~~ p, s  \geq 1 ,
\nonumber \\
&\sim& [p+1; q-1 , r-1; s+1] , ~~ \mbox{for} ~~ q, r  \geq 1 .
\label{equ-rel}
\end{eqnarray}
{}From this observation, we find that the number of equivalence classes 
is given by the difference between the number of diagonal pairs $(P_0, P_1)$ and the number of 
equivalence relations among those pairs and it is
$(N+1)^2$ for $SU(N)$ gauge theories on $S^1/Z_2$.

The vacuum energy density is essentially evaluated by using the value 
at the minimum of $V_{\rm eff}$.
The $V_{\rm eff}$ is a function of the background configuration
of gauge field $A^{\rm bg}_M$, some numbers which specify BCs and numbers of species with definite $Z_2$ parities,
\begin{eqnarray}
V_{\rm eff} = V_{\rm eff}(A^{\rm bg}_M; p, q, r, \beta) ,
\label{Veff}
\end{eqnarray}
where $\beta$ is a parameter related to the soft SUSY breaking due to the Scherk-Schwarz mechanism\cite{S&S}.
We can calculate the one-loop effective potential using the generic formula:
\begin{eqnarray}
V_{\rm eff} = \sum \mp \frac{i}{2} {\rm Tr}~{\rm ln} D_M(A^{\rm bg})D^M(A^{\rm bg})
= \sum \mp i \int \frac{d^4p}{(2\pi)^4} \frac{1}{2\pi R} \sum_{n \in Z} \ln(-p^2 + M_n^2 - i\varepsilon) ,
\label{Veff-formula}
\end{eqnarray}
where the sums extend over all degrees of freedoms of bulk fields, the sign is negative (positive) for bosons
(FP ghosts and fermions) and $M_n$ are masses.
(For the explicit formula in the simple case with the vanishing VEV of $A_y$, see Ref.\cite{HHK}.)  

The BCs including the Scherk-Schwarz SUSY breaking mechanism are given for gauge multiplet and a Higgs hypermultiplet on $S^1/Z_2$ as
\begin{eqnarray}
&~& \left(\begin{array}{c}
V \\
\Sigma 
\end{array}\right)(x, -y)
= P_0 
\left(\begin{array}{c}
V \\
\Sigma 
\end{array}\right)(x, y)  P_0^\dagger ,
\nonumber \\
&~& A_{\mu}(x, 2\pi R-y)= P_1  A_{\mu}(x, y)  P_1^\dagger , ~~
A_y(x, 2 \pi R-y)= -P_1  A_y(x, y)  P_1^\dagger , 
\nonumber \\
&~& \left(\begin{array}{c}
\lambda_1 \\
\lambda_2
\end{array}\right) (x, 2\pi R-y)
= e^{-2\pi i \beta \tau_2} 
P_1 \left(\begin{array}{c}
\lambda_1 \\
-\lambda_2
\end{array}\right) (x, y) P_1^\dagger , ~~
\sigma(x, 2\pi R-y)= -P_1 \sigma(x, y) P_1^\dagger , 
\nonumber \\ 
&~&  A_M(x, y+2 \pi R)= U A_M(x, y) U^\dagger , ~~
\left(\begin{array}{c}
\lambda_1 \\
\lambda_2
\end{array}\right)(x, y+ 2 \pi R)
= e^{-2\pi i \beta \tau_2} U 
\left(\begin{array}{c}
\lambda_1 \\
\lambda_2
\end{array}\right)(x,y) U^\dagger ,
\nonumber \\
&~& \sigma(x, y+2 \pi R)= U \sigma(x,y) U^\dagger ,
\label{SBC1} \\
&~& \left(
\begin{array}{c}
h \\
h^{c\dagger}
\end{array}
\right) (x, -y)
= \eta_0 T_{\cal H}[P_0]
\left(
\begin{array}{c}
h \\
h^{c\dagger}
\end{array}
\right)(x, y) , ~~
\left(
\begin{array}{c}
h \\
h^{c\dagger}
\end{array}
\right)(x, 2\pi R-y)
= e^{-2\pi i \beta \tau_2} ~ \eta_1 T_{\cal H}[P_1]
\left(
\begin{array}{c}
h \\
h^{c\dagger}
\end{array}
\right)(x, y) ,
\nonumber \\
&~& \left(
\begin{array}{c}
h \\
h^{c\dagger}
\end{array}
\right)(x, y + 2\pi R)
= e^{-2\pi i \beta \tau_2} ~ \eta_0 \eta_1 T_{\cal H}[U]
\left(
\begin{array}{c}
h \\
h^{c\dagger}
\end{array}
\right)(x, y) ,
\nonumber \\
&~& \left(
\begin{array}{c}
\tilde{h} \\
\tilde{h}^{c\dagger}
\end{array}
\right)(x, -y)
= \eta_0 T_{\cal H}[P_0]
\left(
\begin{array}{c}
\tilde{h} \\
\tilde{h}^{c\dagger}
\end{array}
\right)(x, y) , ~~
\left(
\begin{array}{c}
\tilde{h} \\
\tilde{h}^{c\dagger}
\end{array}
\right)(x, 2\pi R-y)
= \eta_1 T_{\cal H}[P_1]
\left(
\begin{array}{c}
\tilde{h} \\
\tilde{h}^{c\dagger}
\end{array}
\right)(x, y) ,
\nonumber \\
&~& \left(
\begin{array}{c}
\tilde{h} \\
\tilde{h}^{c\dagger}
\end{array}
\right)(x, y + 2\pi R)
= \eta_0 \eta_1 T_{\cal H}[U]
\left(
\begin{array}{c}
\tilde{h} \\
\tilde{h}^{c\dagger}
\end{array}
\right)(x, y) ,
\label{SBC2}
\end{eqnarray}
where $\beta/R \le O(10^3)$GeV from the phenomenological viewpoint.

Owing to the SUSY, the one-loop effective potential takes a finite value at the minimum 
even after the Scherk-Schwarz SUSY breaking mechanism works.
Hence, we can compare with the vacuum energy density among theories that belong to
different equivalence classes if it were allowed.
We find that it is difficult to realize $[p; q, r; s] =[2; 0, 0; 3]$, which is equivalent to (\ref{GUT-parity}),
as the prefered equivalence class.
Further, irrespective of matter content in the bulk, there remains the degeneracy,
because the $V_{\rm eff}$ is a function of $q+r$ only and a unique $[p; q, r; s]$ is not selected. 
We need to find a mechanism to lift the degeneracy.

Many people doubt if the comparison among gauge-inequivalent theories 
is meaningful or not.
We hope that it can make sense in the situation that 
a fundamental theory has a bigger symmetry and BCs are dynamically determined.

\section{Orbifold family unification}

\subsection{Preparations}

The grand unification is attractive, because it
offers the unification of forces and (partial) unification of quarks and leptons\cite{G&G}.
In the $SU(5)$ GUT, two kinds of representations $\bar{\bf{5}}$ and $\bf{10}$ are introduced for each family.
In the $SO(10)$ GUT, one multiplet ${\bf{16}}$ for each family.
Here the following question comes to mind.
\begin{flushleft}
\begin{tabular}{|l|}
\hline
~~~ {\bf Basic question}\\
Is the unification of families possible or not?\\ \hline
\end{tabular}
\end{flushleft}
The family unification scenarios on a basis of larger symmetry groups are proposed\cite{R,GF,K&Y}.
Cosmological and astrophysical implications are studied on the breakdown of family symmetry\cite{Khlopov}.

In the 4-dimensional Minkowski space-time, we encounter difficulty in the (complete) family unification
because of extra fields such as `mirror particles' existing in the higher-dimensional representation.
Here, the mirror particles are particles with opposite quantum numbers under $G_{\rm SM}$.
If the idea of the (complete) family unification is to be realized in nature, extra particles must disappear 
from the low-energy spectrum around the weak scale.
Several interesting mechanisms have been proposed to get rid of the unwelcomed particles.
One is to confine extra particles at a high-energy scale by some strong interaction~\cite{GR&S}.
Another possibility is to reduce symmetries and substances using extra dimensions, 
as originally discussed in superstring theory~\cite{CHS&W,Orbifold}.
4-dimensional chiral fermions originate through the dimensional reduction 
where some of the zero modes are projected out 
by orbifolding, i.e., by non-trivial BCs concerning the extra dimensions on bulk fields.
Hence we expect that all the extra particles 
plaguing the family unification models can be eliminated from the spectrum 
in the framework of orbifold GUTs and that the idea of the (complete) family unification can be realized.\footnote{
The possibility that one might realize the complete family unification utilizing an orbifold 
is also suggested in Ref.~\cite{BB&K} in a different context.
In Ref.~\cite{Watari:2002tf}, three families are derived from a combination of a bulk gauge multiplet and a few brane fields.
In Ref.~\cite{Chaichian:2001fs}, they are realized as composite fields.}

In 5-dimensional space-time, bulk fields with arbitrary representations are allowed in the first place,
because there are no local anomalies.
There is a possibility or scenario that three families survive from a few hypermultiplets after orbifolding.
On the other hand, the theory can be anomalous on the 4-dimensional boundaries with the appearance of chiral fermions.
Such anomalies must be cancelled in the 4-dimensional effective theory by the contribution of the brane chiral fermions
and/or counterterms, such as the Chern-Simons term~\cite{Anomaly}.
Hence there is a possibility that three families originates from a few hypermultiplets plus some brane fields.
Bearing this observation in mind,
let us derive three families from a hypermultiplet with a large representation 
and, if neccesarry, brane fields in the framework of an $SU(N)$ gauge theory on $M^4 \times (S^1/Z_2)$.
We would like to stress a difference between our work and the previous ones.
Various unification scenarios were, in most case, studied using the gauge supermultiplet.
We use a hypermultiplet in place of the gauge supermultiplet.

Here I clear problems up by listing two questions as follows.
\begin{flushleft}
\begin{tabular}{|l|}
\hline
~~~ {\bf Two questions}\\
1. Are three families in $SU(5)$ GUT derived from a bulk field with the $k$-th anti-symmetric tensor representation\\
~~ of $SU(N)$ or not after the $Z_2$ orbifold breaking?\\
2. Are three families in the SM are derived from a bulk field with the $k$-th anti-symmetric tensor representation of\\
~~ $SU(N)$ or not after the $Z_2$ orbifold breaking?\\ 
\hline
\end{tabular}
\end{flushleft}

We prepare the basic building blocks for our argument\cite{KK&O}.
For simplicity, we consider the symmetry breaking pattern 
$SU(N) \to  SU(p) \times SU(q) \times SU(r) \times SU(s) \times U(1)^\nu$,
which is induced by the representation matrices of the $Z_2$ parities:
\begin{eqnarray}
&~& P_0 = \mbox{diag}(\overbrace{+1, \dots, +1, +1, \dots, +1, -1, \dots, -1, -1, \dots, -1}^N) , 
\label{P0} \\
&~& P_1 = \mbox{diag}(\underbrace{+1, \dots, +1}_{p}, \underbrace{-1, \dots, -1}_{q}, \underbrace{+1, \dots, +1}_{r}, 
 \underbrace{-1, \dots, -1}_{s=N-p-q-r}) ,
\label{P1}
\end{eqnarray}
where ``$SU(1)$'' unconventionally stands for $U(1)$, $SU(0)$ means nothing
and $\nu = 3 - \kappa$ where $\kappa$ is the number of zero or one in $p$, $q$, $r$ and $s$.

After the breakdown of~$SU(N)$, the rank-$k$ completely antisymmetric tensor representation~$[N, k]$, 
whose dimension is~${}_{N}C_{k}$,
is decomposed into a sum of multiplets of the subgroup $SU(p) \times SU(q) \times SU(r) \times SU(s)$ as
\begin{eqnarray}
[N, k] = \sum_{l_1 =0}^{k} \sum_{l_2 = 0}^{k-l_1} \sum_{l_3 = 0}^{k-l_1-l_2}  
\left({}_{p}C_{l_1}, {}_{q}C_{l_2}, {}_{r}C_{l_3}, {}_{s}C_{l_4}\right) ,
\label{Nk}
\end{eqnarray}
where $l_1$, $l_2$ and $l_3$ are integers, $l_4=k-l_1-l_2-l_3$ and our notation is 
such that ${}_{n}C_{l} = 0$ for $l > n$ and $l < 0$.
Here and hereafter, we use ${}_{n}C_{l}$ instead of $[n, l]$ in many cases.
(We sometimes use the ordinary notation for representations too, 
e.g., ${\bf{5}}$ and ${\overline{\bf{5}}}$ in place of ${}_{5}C_{1}$ and ${}_{5}C_{4}$.) 

The $[N, k]$ is constructed by the antisymmetrization of $k$-ple product of the fundamental representation $N = [N, 1]$:
\begin{eqnarray}
[N, k] = (N \times \dots \times N)_a .
\label{N*...*N} 
\end{eqnarray}
We define the intrinsic $Z_2$ and $Z_2'$ parities $\eta_{[N,k]}$ and $\eta'_{[N,k]}$, respectively, such that
\begin{eqnarray}
(N \times \dots \times N)_a \to \eta_{[N,k]} (P_0 N \times \dots \times P_0 N)_a , ~~
(N \times \dots \times N)_a \to \eta'_{[N,k]} (P_1 N \times \dots \times P_1 N)_a .
\label{etaNk} 
\end{eqnarray}
By definition, $\eta_{[N,k]}$ and $\eta'_{[N,k]}$ each takes the value $+1$ or $-1$.
The $Z_2$ parities of the representation~$({}_{p}C_{l_1}, {}_{q}C_{l_2},$ ${}_{r}C_{l_3},$ ${}_{s}C_{l_4})$ are given by
\begin{eqnarray}
\mathcal{P}_0 = (-1)^{l_3+l_4} \eta_{[N,k]} = (-1)^{l_1+l_2} (-1)^k \eta_{[N,k]} , ~~
\mathcal{P}_1 = (-1)^{l_2+l_4} \eta'_{[N,k]} = (-1)^{l_1+l_3} (-1)^k \eta'_{[N,k]} .
\label{Z2}
\end{eqnarray}

A fermion with spin $1/2$ in 5 dimensions is regarded as a Dirac fermion 
or a pair of Weyl fermions with opposite chiralities in 4 dimensions.
The representations of each Weyl fermion are decomposed as,
\begin{eqnarray}
[N, k]_L = \sum_{l_1 =0}^{k} \sum_{l_2 = 0}^{k-l_1} \sum_{l_3 = 0}^{k-l_1-l_2}  
\left({}_{p}C_{l_1}, {}_{q}C_{l_2}, {}_{r}C_{l_3}, {}_{s}C_{l_4}\right)_L , ~~
[N, k]_R = \sum_{l_1 =0}^{k} \sum_{l_2 = 0}^{k-l_1} \sum_{l_3 = 0}^{k-l_1-l_2}  
\left({}_{p}C_{l_1}, {}_{q}C_{l_2}, {}_{r}C_{l_3}, {}_{s}C_{l_4}\right)_R ,
\label{NkLR}
\end{eqnarray}
where the subscript $L$ ($R$) represents left-handedness (right-handedness) for Weyl fermions.
The $Z_2$ parities of the representation~$({}_{p}C_{l_1}, {}_{q}C_{l_2},$ ${}_{r}C_{l_3},$ ${}_{s}C_{l_4})_L$ are given by
\begin{eqnarray}
\mathcal{P}_0 = (-1)^{l_1+l_2} (-1)^k \eta_{[N,k]_L} , ~~
\mathcal{P}_1 = (-1)^{l_1+l_3} (-1)^k \eta'_{[N,k]_L} .
\label{Z2L}
\end{eqnarray}
In the same way, the $Z_2$ parities of the representations~$({}_{p}C_{l_1}, {}_{q}C_{l_2},$ ${}_{r}C_{l_3},$ ${}_{s}C_{l_4})_R$ 
are given by
\begin{eqnarray}
\mathcal{P}_0 = (-1)^{l_1+l_2} (-1)^k \eta_{[N,k]_R} , ~~
\mathcal{P}_1 = (-1)^{l_1+l_3} (-1)^k \eta'_{[N,k]_R} .
\label{Z2R}
\end{eqnarray}
The $({}_{p}C_{l_1}, {}_{q}C_{l_2},$ ${}_{r}C_{l_3},$ ${}_{s}C_{l_4})_L$ and 
$({}_{p}C_{l_1}, {}_{q}C_{l_2},$ ${}_{r}C_{l_3},$ ${}_{s}C_{l_4})_R$ 
should have opposite $Z_2$ parities each other, $\eta_{[N,k]_R} = - \eta_{[N,k]_L}$ and  $\eta'_{[N,k]_R} = - \eta'_{[N,k]_L}$, 
from the requirement that the kinetic term should be invariant under the $Z_2$ parity transformation.
The $Z_2$ transformation property for fermions is written down by
\begin{eqnarray}
(N \times \dots \times N)_a \to - \eta_{[N,k]_L} \gamma_5 (P_0 N \times \dots \times P_0 N)_a , ~~
(N \times \dots \times N)_a \to - \eta'_{[N,k]_L} \gamma_5 (P_1 N \times \dots \times P_1 N)_a ,
\label{etaNkfermion} 
\end{eqnarray}
where $\gamma_5 \psi_L = -\psi_L$ and $\gamma_5 \psi_R = +\psi_R$. 
Hereafter we denote $\eta_{[N,k]_L}$ and $\eta'_{[N,k]_L}$ as $\eta_k$ and $\eta'_k$, respectively.
Both left-handed and right-handed Weyl fermions having even $Z_2$ parities, $\mathcal{P}_0 = \mathcal{P}_1 = +1$, 
compose chiral fermions in the SM.

In SUSY models, the hypermultiplet is the fundamental quantity concerning bulk matter fields in 5 dimensions.
The hypermultiplet is equivalent to a pair of chiral multiplets 
with opposite gauge quantum numbers in 4 dimensions.
The chiral multiplet with the representation $[N,N-k]$, which is a conjugate of $[N,k]$, 
contains a left-handed Weyl fermion with $[N,N-k]_L$.
This Weyl fermion is regarded as a right-handed one with $[N,k]_R$ by using the charge conjugation.
Hence our analysis works on SUSY models as well as non-SUSY ones.

\subsection{Family unification in $SU(N)\to SU(5)$}

We study the gauge symmetry breaking pattern 
$SU(N) \to  SU(5) \times SU(q) \times SU(r) \times SU(s) \times U(1)^\nu$,
which is realized with $Z_2$ parity assignment
\begin{eqnarray}
\hspace{-1.3cm}&~& P_0 = \mbox{diag}(+1, +1, +1, +1, +1, +1, \dots, +1, -1, \dots, -1, -1, \dots, -1) , 
\label{P0-SU5} \\
\hspace{-1.3cm}&~& P_1 = \mbox{diag}(+1, +1, +1, +1, +1, \underbrace{-1, \dots, -1}_{q}, 
\underbrace{+1, \dots, +1}_{r}, \underbrace{-1, \dots, -1}_{s}) ,
\label{P1-SU5}
\end{eqnarray}
where $s = N-5-q-r$.
After the breakdown of $SU(N)$, the $[N, k]$ is decomposed into
a sum of multiplets of the subgroup $SU(5) \times SU(q) \times SU(r) \times SU(s)$ as
\begin{eqnarray}
[N, k] = \sum_{l_1 =0}^{k} \sum_{l_2 = 0}^{k-l_1} \sum_{l_3 = 0}^{k-l_1-l_2}  
\left({}_{5}C_{l_1}, {}_{q}C_{l_2}, {}_{r}C_{l_3}, {}_{s}C_{l_4}\right) .
\label{Nk}
\end{eqnarray}
As mentioned before, 
 ${{}_{5}C_{0}}$, ${{}_{5}C_{1}}$, ${{}_{5}C_{2}}$, ${{}_{5}C_{3}}$, ${{}_{5}C_{4}}$
and ${{}_{5}C_{5}}$ stand for representations ${\bf{1}}$, ${\bf{5}}$, ${\bf{10}}$, 
 ${\overline{\bf{10}}}$, ${\overline{\bf{5}}}$ and ${\overline{\bf{1}}}$.\footnote{
We denote the $SU(5)$ singlet relating to ${{}_{5}C_{5}}$ as ${\overline{\bf{1}}}$, for convenience sake,
to avoid the confusion over singlets.}

Utilizing the survival hypothesis and the equivalence of $({\bf{5}}_R)^c$ and $(\overline{\bf{10}}_R)^c$
with $\overline{\bf{5}}_L$ and ${\bf{10}}_L$, respectively,%
\footnote{
As usual, $({\bf{5}}_R)^c$ and $(\overline{\bf{10}}_R)^c$ represent 
the charge conjugate of ${\bf{5}}_R$ and $\overline{\bf{10}}_R$, respectively. 
Note that $({\bf{5}}_R)^c$ and $(\overline{\bf{10}}_R)^c$ transform as 
left-handed Weyl fermions under the 4-dimensional Lorentz transformations.}
we write the numbers of $\overline{\bf 5}$ and ${\bf{10}}$ representations for left-handed Weyl fermions as
\begin{eqnarray}
n_{\bar{5}} \equiv \sharp{\overline{\bf 5}}_L  - \sharp{\bf 5}_L 
  + \sharp{\bf 5}_R  - \sharp{\overline{\bf 5}}_R , ~~  
n_{10} \equiv \sharp{\bf 10}_L  - \sharp{\overline{\bf 10}}_L 
  + \sharp{\overline{\bf 10}}_R  - \sharp{\bf 10}_R  , 
\label{n510-def}
\end{eqnarray}
where $\sharp$ represents the number of each multiplet. 
When we take $\left((-1)^{k} \eta_{k}, (-1)^{k} \eta'_{k}\right)=\left(+1, +1\right)$, $n_{\bar{5}}$ and $n_{10}$ are given by
\begin{eqnarray}
&~& n_{\bar{5}}
 = \sum_{l_1 = 1, 4} \sum_{l_2 = 0, 2, \dots} \sum_{l_3 = 0, 2, \dots}
  {}_{q}C_{l_2} \cdot {}_{r}C_{l_3} \cdot {}_{s}C_{l_4} 
- \sum_{l_1 = 1, 4} \sum_{l_2 = 1, 3, \dots} \sum_{l_3 = 1, 3, \dots}
  {}_{q}C_{l_2} \cdot {}_{r}C_{l_3} \cdot {}_{s}C_{l_4} \equiv n_{\bar{5},k}^{(++)} ,  
\label{nbar5++}\\
&~& n_{10}
 = \sum_{l_1 = 2, 3} \sum_{l_2 = 0, 2, \dots} \sum_{l_3 = 0, 2, \dots}
  {}_{q}C_{l_2} \cdot {}_{r}C_{l_3} \cdot {}_{s}C_{l_4} 
- \sum_{l_1 = 2, 3} \sum_{l_2 = 1, 3, \dots} \sum_{l_3 = 1, 3, \dots}
  {}_{q}C_{l_2} \cdot {}_{r}C_{l_3} \cdot {}_{s}C_{l_4} \equiv n_{10,k}^{(++)} .
\label{n10++}
\end{eqnarray}
When we take $\left((-1)^{k} \eta_{k} , (-1)^{k} \eta'_{k}\right)=\left(+1, -1\right)$, $n_{\bar{5}}$ and $n_{10}$ are given by
\begin{eqnarray}
&~& n_{\bar{5}}
 = \sum_{l_1 = 1, 4} \sum_{l_2 = 0, 2, \dots} \sum_{l_3 = 1, 3, \dots}
  {}_{q}C_{l_2} \cdot {}_{r}C_{l_3} \cdot {}_{s}C_{l_4} 
- \sum_{l_1 = 1, 4} \sum_{l_2 = 1, 3, \dots} \sum_{l_3 = 0, 2, \dots}
  {}_{q}C_{l_2} \cdot {}_{r}C_{l_3} \cdot {}_{s}C_{l_4} \equiv n_{\bar{5},k}^{(+-)} ,  
\label{nbar5+-}\\
&~& n_{10}
 = \sum_{l_1 = 2, 3} \sum_{l_2 = 0, 2, \dots} \sum_{l_3 = 1, 3, \dots}
  {}_{q}C_{l_2} \cdot {}_{r}C_{l_3} \cdot {}_{s}C_{l_4} 
- \sum_{l_1 = 2, 3} \sum_{l_2 = 1, 3, \dots} \sum_{l_3 = 0, 2, \dots}
  {}_{q}C_{l_2} \cdot {}_{r}C_{l_3} \cdot {}_{s}C_{l_4} \equiv n_{10,k}^{(+-)} .
\label{n10+-}
\end{eqnarray}
In the same way, we can derive $n_{\bar{5}} = -n_{\bar{5},k}^{(+-)}$ and $n_{10} = - n_{10,k}^{(+-)}$ 
for $\left((-1)^{k} \eta_{k}, (-1)^{k} \eta'_{k}\right)=\left(-1, +1\right)$
and $n_{\bar{5}} = -n_{\bar{5},k}^{(++)}$ and $n_{10} = - n_{10,k}^{(++)}$ 
for $\left((-1)^{k} \eta_{k}, (-1)^{k} \eta'_{k}\right)=\left(-1, -1\right)$.

\begin{flushleft}
\begin{tabular}{|l|}
\hline
~~~ {\bf Answer to the first question}\\
There are many possibilities to derive three families $n_{\bar{5}} = n_{10} = 3$.\\ \hline
\end{tabular}
\end{flushleft}
The representations and BCs derived three families up to $SU(15)$ are listed in Table \ref{t1}.
\begin{table}
\vspace{-0.6cm}
\caption{Representations and BCs derived three families up to $SU(15)$}
\label{t1}
\begin{tabular}{c|c|c|c} \hline
 Representation & $[p; q, r; s]$ & $(-1)^{k} \eta_{k}$ & $(-1)^k \eta'_{k}$ \\ \hline
 $[9,3]$ & $[5; 0, 3; 1]$ & $+1$ & $-1$ \\ 
 $[9,3]$ & $[5; 3, 0; 1]$ & $-1$ & $+1$ \\ 
 $[9,6]$ & $[5; 0, 3; 1]$ & $+1$ & $+1$ \\ 
 $[9,6]$ & $[5; 3, 0; 1]$ & $+1$ & $+1$ \\ \hline
 $[11,3]$ & $[5; 1, 4; 1]$ & $+1$ & $-1$ \\ 
 $[11,3]$ & $[5; 4, 1; 1]$ & $-1$ & $+1$ \\ 
 $[11,4]$ & $[5; 1, 4; 1]$ & $+1$ & $+1$ \\ 
 $[11,4]$ & $[5; 4, 1; 1]$ & $+1$ & $+1$ \\ 
 $[11,7]$ & $[5; 1, 4; 1]$ & $-1$ & $+1$ \\ 
 $[11,7]$ & $[5; 4, 1; 1]$ & $+1$ & $-1$ \\ 
 $[11,8]$ & $[5; 1, 4; 1]$ & $-1$ & $-1$ \\ 
 $[11,8]$ & $[5; 4, 1; 1]$ & $-1$ & $-1$ \\ \hline
 $[12,3]$ & $[5; 1, 4; 2]$ & $+1$ & $+1$ \\ 
 $[12,3]$ & $[5; 4, 1; 2]$ & $+1$ & $+1$ \\ 
 $[12,9]$ & $[5; 1, 4; 2]$ & $-1$ & $+1$ \\ 
 $[12,9]$ & $[5; 4, 1; 2]$ & $+1$ & $-1$ \\ \hline
 $[13,3]$ & $[5; 2, 5; 1]$ & $+1$ & $-1$ \\ 
 $[13,3]$ & $[5; 5, 2; 1]$ & $-1$ & $+1$ \\ 
 $[13,10]$ & $[5; 2, 5; 1]$ & $+1$ & $+1$ \\ 
 $[13,10]$ & $[5; 5, 2; 1]$ & $+1$ & $+1$ \\ \hline
 $[14,4]$ & $[5; 4, 4; 1]$ & $-1$ & $-1$ \\ 
 $[14,10]$ & $[5; 2, 6; 1]$ & $+1$ & $+1$ \\ 
 $[14,10]$ & $[5; 4, 4; 1]$ & $-1$ & $-1$ \\ 
 $[14,10]$ & $[5; 6, 2; 1]$ & $+1$ & $+1$ \\ \hline
 $[15,3]$ & $[5; 3, 6; 1]$ & $+1$ & $-1$ \\ 
 $[15,3]$ & $[5; 6, 3; 1]$ & $-1$ & $+1$ \\ 
 $[15,4]$ & $[5; 4, 5; 1]$ & $-1$ & $-1$ \\ 
 $[15,4]$ & $[5; 5, 4; 1]$ & $-1$ & $-1$ \\ 
 $[15,5]$ & $[5; 4, 5; 1]$ & $-1$ & $+1$ \\ 
 $[15,5]$ & $[5; 5, 4; 1]$ & $+1$ & $-1$ \\ 
 $[15,10]$ & $[5; 4, 5; 1]$ & $-1$ & $-1$ \\ 
 $[15,10]$ & $[5; 5, 4; 1]$ & $-1$ & $-1$ \\ 
 $[15,11]$ & $[5; 3, 6; 1]$ & $+1$ & $-1$ \\ 
 $[15,11]$ & $[5; 4, 5; 1]$ & $-1$ & $+1$ \\ 
 $[15,11]$ & $[5; 5, 4; 1]$ & $+1$ & $-1$ \\ 
 $[15,11]$ & $[5; 6, 3; 1]$ & $-1$ & $+1$ \\ \hline
\end{tabular}
\end{table}

\subsection{Family unification in $SU(N)\to G_{\rm SM}$}

We study the gauge symmetry breaking pattern, $SU(N) \to SU(3) \times SU(2) \times SU(r) \times SU(s) \times U(1)^n$,
which is realized by the $Z_2$ parity assignment
\begin{eqnarray}
&~& P_0 = \mbox{diag}(+1, +1, +1, +1, +1, -1, \dots, -1, -1, \dots, -1) , 
\label{P0-SM} \\
&~& P_1 = \mbox{diag}(+1, +1, +1, -1, -1, \underbrace{+1, \dots, +1}_{r}, \underbrace{-1, \dots, -1}_{s}) ,
\label{P1-SM}
\end{eqnarray}
where $s = N-5-r$ and $N \ge 6$.
After the breakdown of $SU(N)$, the $[N, k]$ is decomposed into
a sum of multiplets of the subgroup $SU(3) \times SU(2) \times SU(r) \times SU(s)$ as
\begin{eqnarray}
[N, k] = \sum_{l_1 =0}^{k} \sum_{l_2 = 0}^{k-l_1} \sum_{l_3 = 0}^{k-l_1-l_2}  
\left({}_{3}C_{l_1}, {}_{2}C_{l_2}, {}_{r}C_{l_3}, {}_{s}C_{l_4}\right) .
\label{Nk}
\end{eqnarray}
We list the $U(1)$ charges for representations of the subgroups in Table \ref{t2}.
\begin{table}
\caption{The $U(1)$ charges for representations of fermions}
\label{t2}
\begin{tabular}{c|c|c|c|c} \hline
species & representation  & $U(1)_1$ & $U(1)_2$ & $U(1)_3$ \\ \hline
$(\nu_{R})^c$, $\hat{\nu}_{R}$ & $\left({}_{3}C_{0}, {}_{2}C_{0}, {}_{r}C_{l_3}, {}_{s}C_{k-l_3}\right)$ 
& $0$ & $(N-5)l_3 - rk$ & $-5k$ \\ \hline
$(d'_{R})^c$, $d_{R}$ & $\left({}_{3}C_{1}, {}_{2}C_{0}, {}_{r}C_{l_3}, {}_{s}C_{k-l_3-1}\right)$ & $-2$ 
& $(N-5)l_3 - r(k-1)$ & $N-5k$ \\
$l'_{L}$, $(l_{L})^c$ & $\left({}_{3}C_{0}, {}_{2}C_{1}, {}_{r}C_{l_3}, {}_{s}C_{k-l_3-1}\right)$ & $3$ 
& $(N-5)l_3 - r(k-1)$ & $N-5k$ \\ \hline
$(u_{R})^c$, $u'_{R}$ & $\left({}_{3}C_{2}, {}_{2}C_{0}, {}_{r}C_{l_3}, {}_{s}C_{k-l_3-2}\right)$ & $-4$ 
& $(N-5)l_3 - r(k-2)$ & $2N-5k$ \\
$(e_{R})^c$, $e'_{R}$ & $\left({}_{3}C_{0}, {}_{2}C_{2}, {}_{r}C_{l_3}, {}_{s}C_{k-l_3-2}\right)$ & $6$ 
& $(N-5)l_3 - r(k-2)$ & $2N-5k$ \\
$q_{L}$, $(q'_{L})^c$ & $\left({}_{3}C_{1}, {}_{2}C_{1}, {}_{r}C_{l_3}, {}_{s}C_{k-l_3-2}\right)$ & $1$ 
& $(N-5)l_3 - r(k-2)$ & $2N-5k$ \\ \hline
$(e'_{R})^c$, $e_{R}$ & $\left({}_{3}C_{3}, {}_{2}C_{0}, {}_{r}C_{l_3}, {}_{s}C_{k-l_3-3}\right)$ & $-6$ 
& $(N-5)l_3 - r(k-3)$ & $3N-5k$ \\
$(u'_{R})^c$, $u_{R}$ & $\left({}_{3}C_{1}, {}_{2}C_{2}, {}_{r}C_{l_3}, {}_{s}C_{k-l_3-3}\right)$ & $4$ 
& $(N-5)l_3 - r(k-3)$ & $3N-5k$ \\
$q'_{L}$, $(q_{L})^c$ & $\left({}_{3}C_{2}, {}_{2}C_{1}, {}_{r}C_{l_3}, {}_{s}C_{k-l_3-3}\right)$ & $-1$ 
& $(N-5)l_3 - r(k-3)$ & $3N-5k$ \\ \hline
$l_{L}$, $(l'_{L})^c$ & $\left({}_{3}C_{3}, {}_{2}C_{1}, {}_{r}C_{l_3}, {}_{s}C_{k-l_3-4}\right)$ & $-3$ 
& $(N-5)l_3 - r(k-4)$ & $4N-5k$ \\
$(d_{R})^c$, $d'_{R}$ & $\left({}_{3}C_{2}, {}_{2}C_{2}, {}_{r}C_{l_3}, {}_{s}C_{k-l_3-4}\right)$ & $2$ 
& $(N-5)l_3 - r(k-4)$ & $4N-5k$ \\ \hline
$(\hat{\nu}_{R})^c$, $\nu_{R}$ & $\left({}_{3}C_{3}, {}_{2}C_{2}, {}_{r}C_{l_3}, {}_{s}C_{k-l_3-5}\right)$ 
& $0$ & $(N-5)l_3 - r(k-5)$ & $5N-5k$ \\ \hline
\end{tabular}
\end{table}
The $U(1)$ charges are those in the subgroups
\begin{eqnarray}
\hspace{-5mm} &~& SU(5) \supset  SU(3) \times SU(2) \times U(1)_1 , \\
\hspace{-5mm} &~& SU(N-5) \supset  SU(r) \times SU(N-5-r) \times U(1)_2 ,~ 
	         SU(N-5-1) \times U(1)_2 ,  \\
\hspace{-5mm} &~& SU(N) \supset  SU(5) \times SU(N-5) \times U(1)_3,
\end{eqnarray}
up to normalization.
We assume that $G_{\rm SM} = SU(3) \times SU(2) \times U(1)_1$, up to normalization of the hypercharge.
Particle species are identified with the SM fermions by the gauge quantum numbers.
Here, we use $(d_{R})^c$, $l_{L}$, $(u_{R})^c$, $(e_{R})^c$ and $q_{L}$ to represent
down-type anti-quark singlets, lepton doublets, up-type anti-quark singlets, positron-type lepton singlets and quark doublets. 
The particles with primes are regarded as mirror particles and believed to have no zero modes.
Each fermion has a definite chirality, e.g. $(d_{R})^c$ is left-handed and $d_{R}$ is right-handed.
Here, the subscript $L$ ($R$) represents left-handedness (right-handedness) for Weyl fermions.
The $(d_R)^c$ represents the charge conjugate of $d_R$ and transforms 
as a left-handed Weyl fermion under the 4-dimensional Lorentz transformation.

We list the $Z_2$ parity assignment for species in Table \ref{t3}.
Note that mirror particles have the $Z_2$ parity $\mathcal{P}_0 = -(-1)^k \eta_{k}$.
Hence {\it all zero modes of mirror particles can be eliminated by the proper choice of $Z_2$ parity
when we take $(-1)^k \eta_{k} = +1$.}
\begin{table}
\caption{The $Z_2$ parity assignment for representations of fermions}
\label{t3}
\begin{tabular}{c|c|c|c} \hline
species & representation  & $\mathcal{P}_0$ & $\mathcal{P}_1$ \\ \hline
$(\nu_{R})^c$ & $\left({}_{3}C_{0}, {}_{2}C_{0}, {}_{r}C_{l_3}, {}_{s}C_{k-l_3}\right)_L$  
 & $(-1)^k \eta_{k}$ & $(-1)^{l_3} (-1)^k \eta'_{k}$ \\ 
$\hat{\nu}_{R}$ & $\left({}_{3}C_{0}, {}_{2}C_{0}, {}_{r}C_{l_3}, {}_{s}C_{k-l_3}\right)_R$  
 & $-(-1)^k \eta_{k}$ & $-(-1)^{l_3} (-1)^k \eta'_{k}$ \\ \hline
$(d'_{R})^c$ & $\left({}_{3}C_{1}, {}_{2}C_{0}, {}_{r}C_{l_3}, {}_{s}C_{k-l_3-1}\right)_L$  
 & $-(-1)^k \eta_{k}$ & $-(-1)^{l_3} (-1)^k \eta'_{k}$ \\
$l'_{L}$ & $\left({}_{3}C_{0}, {}_{2}C_{1}, {}_{r}C_{l_3}, {}_{s}C_{k-l_3-1}\right)_L$  
 & $-(-1)^k \eta_{k}$  & $(-1)^{l_3} (-1)^k \eta'_{k}$ \\ 
$d_{R}$ & $\left({}_{3}C_{1}, {}_{2}C_{0}, {}_{r}C_{l_3}, {}_{s}C_{k-l_3-1}\right)_R$  
 & $(-1)^k \eta_{k}$ & $(-1)^{l_3} (-1)^k \eta'_{k}$ \\
$(l_{L})^c$ & $\left({}_{3}C_{0}, {}_{2}C_{1}, {}_{r}C_{l_3}, {}_{s}C_{k-l_3-1}\right)_R$  
 & $(-1)^k \eta_{k}$  & $-(-1)^{l_3} (-1)^k \eta'_{k}$ \\ \hline
$(u_{R})^c$ & $\left({}_{3}C_{2}, {}_{2}C_{0}, {}_{r}C_{l_3}, {}_{s}C_{k-l_3-2}\right)_L$ 
 & $(-1)^k \eta_{k}$  & $(-1)^{l_3} (-1)^k \eta'_{k}$ \\
$(e_{R})^c$ & $\left({}_{3}C_{0}, {}_{2}C_{2}, {}_{r}C_{l_3}, {}_{s}C_{k-l_3-2}\right)_L$ 
 & $(-1)^k \eta_{k}$  & $(-1)^{l_3} (-1)^k \eta'_{k}$ \\
$q_{L}$ & $\left({}_{3}C_{1}, {}_{2}C_{1}, {}_{r}C_{l_3}, {}_{s}C_{k-l_3-2}\right)_L$ 
 & $(-1)^k \eta_{k}$  & $-(-1)^{l_3} (-1)^k \eta'_{k}$ \\ 
$u'_{R}$ & $\left({}_{3}C_{2}, {}_{2}C_{0}, {}_{r}C_{l_3}, {}_{s}C_{k-l_3-2}\right)_R$ 
 & $-(-1)^k \eta_{k}$  & $-(-1)^{l_3} (-1)^k \eta'_{k}$ \\
$e'_{R}$ & $\left({}_{3}C_{0}, {}_{2}C_{2}, {}_{r}C_{l_3}, {}_{s}C_{k-l_3-2}\right)_R$ 
 & $-(-1)^k \eta_{k}$  & $-(-1)^{l_3} (-1)^k \eta'_{k}$ \\
$(q'_{L})^c$ & $\left({}_{3}C_{1}, {}_{2}C_{1}, {}_{r}C_{l_3}, {}_{s}C_{k-l_3-2}\right)_R$ 
 & $-(-1)^k \eta_{k}$  & $(-1)^{l_3} (-1)^k \eta'_{k}$ \\ \hline
$(e'_{R})^c$ & $\left({}_{3}C_{3}, {}_{2}C_{0}, {}_{r}C_{l_3}, {}_{s}C_{k-l_3-3}\right)_L$ 
 & $-(-1)^k \eta_{k}$  & $-(-1)^{l_3} (-1)^k \eta'_{k}$ \\
$(u'_{R})^c$ & $\left({}_{3}C_{1}, {}_{2}C_{2}, {}_{r}C_{l_3}, {}_{s}C_{k-l_3-3}\right)_L$ 
 & $-(-1)^k \eta_{k}$  & $-(-1)^{l_3} (-1)^k \eta'_{k}$ \\
$q'_{L}$ & $\left({}_{3}C_{2}, {}_{2}C_{1}, {}_{r}C_{l_3}, {}_{s}C_{k-l_3-3}\right)_L$ 
 & $-(-1)^k \eta_{k}$  & $(-1)^{l_3} (-1)^k \eta'_{k}$ \\ 
$e_{R}$ & $\left({}_{3}C_{3}, {}_{2}C_{0}, {}_{r}C_{l_3}, {}_{s}C_{k-l_3-3}\right)_R$ 
 & $(-1)^k \eta_{k}$  & $(-1)^{l_3} (-1)^k \eta'_{k}$ \\
$u_{R}$ & $\left({}_{3}C_{1}, {}_{2}C_{2}, {}_{r}C_{l_3}, {}_{s}C_{k-l_3-3}\right)_R$ 
 & $(-1)^k \eta_{k}$  & $(-1)^{l_3} (-1)^k \eta'_{k}$ \\
$(q_{L})^c$ & $\left({}_{3}C_{2}, {}_{2}C_{1}, {}_{r}C_{l_3}, {}_{s}C_{k-l_3-3}\right)_R$ 
 & $(-1)^k \eta_{k}$  & $-(-1)^{l_3} (-1)^k \eta'_{k}$ \\ \hline
$l_{L}$ & $\left({}_{3}C_{3}, {}_{2}C_{1}, {}_{r}C_{l_3}, {}_{s}C_{k-l_3-4}\right)_L$ 
 & $(-1)^k \eta_{k}$  & $-(-1)^{l_3} (-1)^k \eta'_{k}$ \\
$(d_{R})^c$ & $\left({}_{3}C_{2}, {}_{2}C_{2}, {}_{r}C_{l_3}, {}_{s}C_{k-l_3-4}\right)_L$ 
 & $(-1)^k \eta_{k}$  & $(-1)^{l_3} (-1)^k \eta'_{k}$ \\ 
$(l'_{L})^c$ & $\left({}_{3}C_{3}, {}_{2}C_{1}, {}_{r}C_{l_3}, {}_{s}C_{k-l_3-4}\right)_R$ 
 & $-(-1)^k \eta_{k}$  & $(-1)^{l_3} (-1)^k \eta'_{k}$ \\
$d'_{R}$ & $\left({}_{3}C_{2}, {}_{2}C_{2}, {}_{r}C_{l_3}, {}_{s}C_{k-l_3-4}\right)_R$ 
 & $-(-1)^k \eta_{k}$  & $-(-1)^{l_3} (-1)^k \eta'_{k}$ \\ \hline
$(\hat{\nu}_{R})^c$ & $\left({}_{3}C_{3}, {}_{2}C_{2}, {}_{r}C_{l_3}, {}_{s}C_{k-l_3-5}\right)_L$ 
 & $-(-1)^k \eta_{k}$  & $-(-1)^{l_3} (-1)^k \eta'_{k}$ \\ 
$\nu_{R}$ & $\left({}_{3}C_{3}, {}_{2}C_{2}, {}_{r}C_{l_3}, {}_{s}C_{k-l_3-5}\right)_R$ 
 & $(-1)^k \eta_{k}$  & $(-1)^{l_3} (-1)^k \eta'_{k}$ \\ \hline
\end{tabular}
\end{table}

We write the flavor numbers of $(d_{R})^c$, $l_{L}$, $(u_{R})^c$, $(e_{R})^c$, $q_{L}$ 
and the (heavy) neutrino singlets as $n_{\bar{d}}$, $n_l$, $n_{\bar{u}}$, $n_{\bar{e}}$, $n_q$ and $n_{\bar{\nu}}$.
When we choose $(-1)^k \eta_{k} = (-1)^k \eta'_{k} = +1$, the flavor numbers of the chiral fermions are given by
\begin{eqnarray}
&~& n_{\bar{d}} 
	= \sum_{i = 1, 4} \sum_{l_3 = 0, 2, \dots} {}_{r}C_{l_3} \cdot {}_{N-5-r}C_{k-i-l_3} ,
\label{nd}\\
&~& n_{l}
	= \sum_{i = 1, 4} \sum_{l_3 = 1, 3, \dots} {}_{r}C_{l_3} \cdot {}_{N-5-r}C_{k-i-l_3} ,
\label{nl}\\
&~& n_{\bar{u}} = n_{\bar{e}} 
	= \sum_{i = 2, 3} \sum_{l_3 = 0, 2, \dots} {}_{r}C_{l_3} \cdot {}_{N-5-r}C_{k-i-l_3} ,
\label{ne}\\
&~& n_{q} 
	= \sum_{i = 2, 3} \sum_{l_3 = 1, 3, \dots} {}_{r}C_{l_3} \cdot {}_{N-5-r}C_{k-i-l_3} ,
\label{nq}\\
&~& n_{\bar{\nu}} 
	= \sum_{i = 0, 5} \sum_{l_3 = 0, 2, \dots} {}_{r}C_{l_3} \cdot {}_{N-5-r}C_{k-i-l_3} ,
\label{nnu}
\end{eqnarray}
using the equivalence of the charge conjugation.
When we choose $(-1)^k \eta'_{k} = -1$, we obtain formulae in which $n_l$ is replaced by $n_{\bar{d}}$
and $n_q$ by $n_{\bar{u}}$ ($=n_{\bar{e}}$) in Eqs. (\ref{nd}) - (\ref{nq}).
The total number of (heavy) neutrino singlets is given by
$n_{\bar{\nu}} = \sum_{i = 0, 5} \sum_{l_3 = 1, 3, \dots} {}_{r}C_{l_3} \cdot {}_{N-5-r}C_{k-i-l_3}$
for $(-1)^k \eta'_{k} = -1$.

For arbitrary $N~(\ge 6)$ and $r$, the flavor numbers from $[N, k]$ 
with $((-1)^k \eta_{k},$ $(-1)^k \eta'_{k})$ $= (a, b)$ are
equal to those from $[N, N-k]$ with $((-1)^{N-k} \eta_{N-k}$, $(-1)^{N-k} \eta'_{N-k})$ $= (a, -b)$ if $r$ is odd,
and the flavor numbers from $[N, k]$ 
with $((-1)^k \eta_{k},$ $(-1)^k \eta'_{k})$ $= (a, b)$ are
equal to those from $[N, N-k]$ with $((-1)^{N-k} \eta_{N-k},$ $(-1)^{N-k} \eta'_{N-k})$ $= (a, b)$ if $r$ is even.
We list the flavor number of each chiral fermion derived from $[N, k]$ 
($N = 5, \cdots, 9$ and $k = 1, \cdots, [N/2]$
where $[*]$ stands for Gauss's symbol, i.e., $[N/2] = N/2$ if $N$ is even and $[N/2] = (N-1)/2$ if $N$ is odd) 
in Table \ref{t4}.
In the 8-th column, the numbers in the parenthesis are the flavor numbers of the neutrino singlets for $(-1)^k \eta'_{k} = -1$.
\begin{table}
\caption{The flavor number of each chiral fermion with $(-1)^k \eta_{k} = (-1)^k \eta'_{k} = +1$}
\label{t4}
\begin{tabular}{c|c|c|c|c|c|c|c}
\hline
representation &$(p,q,r,s)$&$n_{\bar{d}}$&$n_{l}$&$n_{\bar{u}}$&$n_{\bar{e}}$&$n_{q}$&$n_{\bar{\nu}}$ 
($n_{\bar{\nu}}$ with $(-1)^k \eta'_{k} = -1$) \\
\hline
$[N,1]$&$(3,2,r,s)$&1&0&0&0&0&$s~(r)$\\
\hline
$[N,2]$&$(3,2,r,s)$&$s$&$r$&1&1&0&${}_rC_2+{}_sC_2$~$(rs)$\\
\hline
$[6,3]$&(3,2,1,0)&0&0&1&1&1&0~(0)\\
\cline{2-8}
&(3,2,0,1)&0&0&2&2&0&0~(0)\\
\hline
&(3,2,2,0)&1&0&1&1&2&0~(0)\\
\cline{2-8}
$[7,3]$&(3,2,1,1)&0&1&2&2&1&0~(0)\\
\cline{2-8}
&(3,2,0,2)&1&0&3&3&0&0~(0)\\
\hline
&(3,2,3,0)&3&0&1&1&3&0~(1)\\
\cline{2-8}
$[8,3]$&(3,2,2,1)&1&2&2&2&2&1~(0)\\
\cline{2-8}
&(3,2,1,2)&1&2&3&3&1&0~(1)\\
\cline{2-8}
&(3,2,0,3)&3&0&4&4&0&1~(0)\\
\hline
&(3,2,3,0)&1&1&3&3&3&0~(0)\\
\cline{2-8}
$[8,4]$&(3,2,2,1)&2&0&2&2&4&0~(0)\\
\cline{2-8}
&(3,2,1,2)&1&1&3&3&3&0~(0)\\
\cline{2-8}
&(3,2,0,3)&2&0&6&6&0&0~(0)\\
\hline
&(3,2,4,0)&6&0&1&1&4&0~(4)\\
\cline{2-8}
&(3,2,3,1)&3&3&2&2&3&3~(1)\\
\cline{2-8}
$[9,3]$&(3,2,2,2)&2&4&3&3&2&2~(2)\\
\cline{2-8}
&(3,2,1,3)&3&3&4&4&1&1~(3)\\
\cline{2-8}
&(3,2,0,4)&6&0&5&5&0&4~(0)\\
\hline
&(3,2,4,0)&1&4&6&6&4&1~(0)\\
\cline{2-8}
&(3,2,3,1)&4&1&4&4&6&0~(1)\\
\cline{2-8}
$[9,4]$&(3,2,2,2)&3&2&4&4&6&1~(0)\\
\cline{2-8}
&(3,2,1,3)&2&3&6&6&4&0~(1)\\
\cline{2-8}
&(3,2,0,4)&5&0&10&10&0&1~(0)\\
\hline
\end{tabular}
\end{table}
\begin{flushleft}
\begin{tabular}{|l|}
\hline
~~~ {\bf Answer to the second question}\\
Any solution satisfying $n_{\bar{d}} = n_l = n_{\bar{u}} = n_{\bar{e}} = n_q = n_{\bar{\nu}} = 3$ has not found.\\ \hline
\end{tabular}
\end{flushleft}

Main subjects left behinds are as follows.
\begin{flushleft}
\begin{tabular}{|l|}
\hline
~~~ {\bf Subjects in Orbifold family Unification}\\
1. To derive orbifold family unification models from a fundamental theory.\\
2. To construct a realistic model.\\ \hline
\end{tabular}
\end{flushleft}

For the first subject, the superstring theory (SST) is a possible candidate of a fundamental theory. 
But it is difficult to derive higher-dimensional representations as massless states of string. 
For the second subject, we must find a mechanism to break extra gauge symmetries
and derive a realistic fermion mass matrices, CKM and MNS matrices.
We must also find model-dependent predictions to distingush among models.
Sum rules among superparticles could be useful for the selection of a realistic model
and we will explain it in the next subsection.

\subsection{Sfermion mass relations}

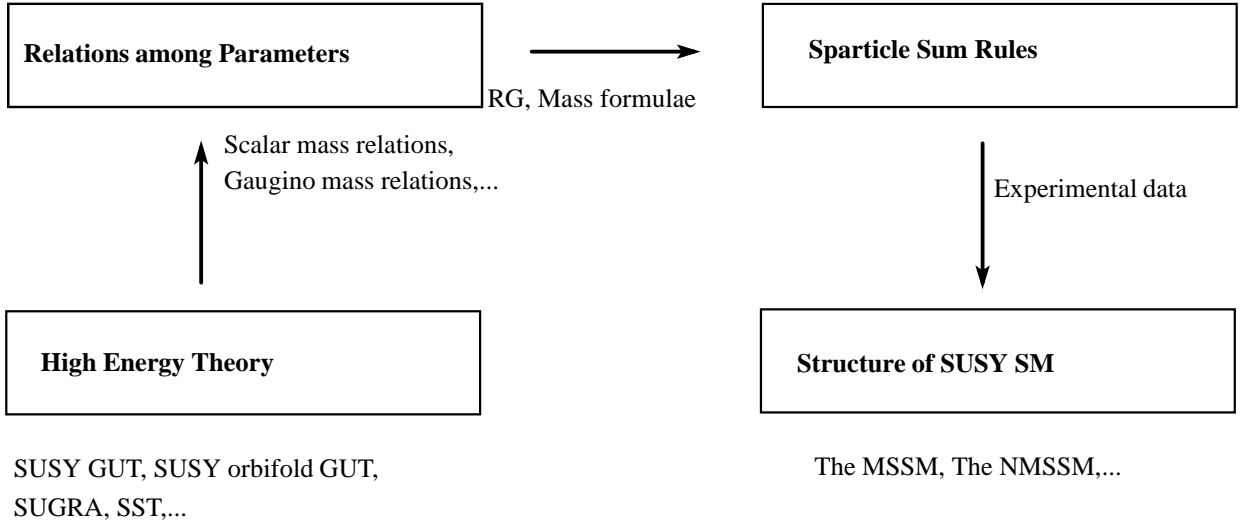
\begin{figure}
\caption{Outline of strategy}
\label{F1}
\unitlength 0.1in
\begin{picture}( 66.0000, 33.3000)( 4.000,-31.1000)
%
\special{pn 13}%
\special{pa 400 380}%
\special{pa 2880 380}%
\special{pa 2880 920}%
\special{pa 400 920}%
\special{pa 400 380}%
\special{fp}%
\put(4.8000,-7.1000){\makebox(0,0)[lb]{\bf{Relations among Parameters}}}%
%
\special{pn 13}%
\special{pa 400 380}%
\special{pa 2880 380}%
\special{pa 2880 920}%
\special{pa 400 920}%
\special{pa 400 380}%
\special{fp}%
%
\special{pn 13}%
\special{pa 390 1990}%
\special{pa 2870 1990}%
\special{pa 2870 2530}%
\special{pa 390 2530}%
\special{pa 390 1990}%
\special{fp}%
%
\special{pn 13}%
\special{pa 4350 380}%
\special{pa 6830 380}%
\special{pa 6830 920}%
\special{pa 4350 920}%
\special{pa 4350 380}%
\special{fp}%
%
\special{pn 13}%
\special{pa 4340 1980}%
\special{pa 6820 1980}%
\special{pa 6820 2520}%
\special{pa 4340 2520}%
\special{pa 4340 1980}%
\special{fp}%
\put(5.7000,-23.3000){\makebox(0,0)[lb]{\bf{High Energy Theory}}}%
\put(45.9000,-6.9000){\makebox(0,0)[lb]{\bf{Sparticle Sum Rules}}}%
\put(45.3000,-23.1000){\makebox(0,0)[lb]{\bf{Structure of SUSY SM}}}%
%
\special{pn 20}%
\special{pa 1410 1840}%
\special{pa 1410 1130}%
\special{fp}%
\special{sh 1}%
\special{pa 1410 1130}%
\special{pa 1390 1198}%
\special{pa 1410 1184}%
\special{pa 1430 1198}%
\special{pa 1410 1130}%
\special{fp}%
\put(15.3000,-11.8000){\makebox(0,0)[lb]{Scalar mass relations,}}%
\put(15.3000,-13.8000){\makebox(0,0)[lb]{Gaugino mass relations,...}}%
\put(4.3000,-28.8000){\makebox(0,0)[lb]{SUSY GUT, SUSY orbifold GUT,}}%
\put(4.3000,-30.8000){\makebox(0,0)[lb]{SUGRA, SST,...}}%
\put(46.2000,-28.7000){\makebox(0,0)[lb]{The MSSM, The NMSSM,...}}%
%
\special{pn 20}%
\special{pa 3140 630}%
\special{pa 3990 630}%
\special{fp}%
\special{sh 1}%
\special{pa 3990 630}%
\special{pa 3924 610}%
\special{pa 3938 630}%
\special{pa 3924 650}%
\special{pa 3990 630}%
\special{fp}%
\put(29.1000,-9.4000){\makebox(0,0)[lb]{RG, Mass formulae}}%
%
\special{pn 20}%
\special{pa 5500 1110}%
\special{pa 5500 1840}%
\special{fp}%
\special{sh 1}%
\special{pa 5500 1840}%
\special{pa 5520 1774}%
\special{pa 5500 1788}%
\special{pa 5480 1774}%
\special{pa 5500 1840}%
\special{fp}%
\put(55.6000,-14.2000){\makebox(0,0)[lb]{Experimental data}}%
\end{picture}%
\end{figure}

First we explain the outline of our strategy according to Figure \ref{F1}.
Let us construct a high energy theory with particular particle contents and (unified gauge) symmetries
including SUSY.
The theory, in general, contains free parameters including unknown quantities related to symmetry breakings, e.g., 
$D$-term contributions to scalar masses\cite{Dterm}.
We can derive specific relations among the soft SUSY breaking parameters at the unification $M_U$
by eliminating free parameters.\footnote{
Scalar mass relations were derived based on various high energy theories, 
e.g., the MSSM\cite{FHK&N,M&R}, 4-dimensional SUSY GUTs\cite{KM&Y,K&T} and 4-dimensional string models\cite{KK&K,K&K}.}
The soft SUSY breaking parameters receive renormalization group (RG) effects, 
and their values at the TeV scale can be calculated by using the RG equations.
After the breakdown of electroweak symmetry, mass formulae of the physical masses are written, 
in terms of parameters in the SUSY SM.
Peculiar sum rules among sparticle masses at the TeV scale are obtained by rewriting
specific relations at $M_U$ in terms of physical masses and parameters. 
In the near future, if superpartners and Higgs bosons were discoverd and their masses 
and interactions were precisely measured,
the presumed high energy theory can be tested by checking whether peculiar sum rules hold or not.

Next, as a simple example, we study the sum rules among the sfermion masses 
that come from the $[8, 3]$ of $SU(8)$ after the orbifold breaking 
$SU(8) \to G_{SM} \times SU(3) \times U(1)_3$, with $(p, q, r, s)=(3, 2, 3, 0)$,
under the assumption that the MSSM holds from the TeV scale
to $M_U$ and that the conventional RG equations of soft SUSY breaking parameters are valid.\footnote{
For detailed descriptions of the methods and derivations, see Ref.\cite{K&Kinami}.}$^,$\footnote{
In Ref.\cite{DFGSS&T}, Dine et al. pointed out that hidden sector interactions
can give rise to sizable effects on the RG evolution of soft SUSY breaking parameters if hidden sector fields
are treated as dynamical.
In Ref.\cite{CR&S}, Cohen et al. derived mass relations among scalar fields by using 
RGEs modified by the hidden dynamics from the GUT scale to an intermediate scale, where auxiliary fields in the hidden sector 
freeze into their VEV.}
After the breakdown of $SU(8)$, the third antisymmetric representation, $[8, 3]$, with ${}_{8}C_{3}$ components
is decomposed into a sum of multiplets of the subgroup $SU(3)_C \times SU(2)_L \times SU(3)$, 
\begin{eqnarray}
[8, 3] = \sum_{l_1 =0}^{3} \sum_{l_2 = 0}^{3-l_1} \left({}_{3}C_{l_1}, {}_{2}C_{l_2}, {}_{3}C_{3-l_1-l_2}\right) .
\label{83}
\end{eqnarray}
The $Z_2$ parity of $\left({}_{3}C_{l_1}, {}_{2}C_{l_2}, {}_{3}C_{3-l_1-l_2}\right)$ is given by
\begin{eqnarray}
\mathcal{P}_0 =  -(-1)^{l_1+l_2} \eta_{3} ,~ \mathcal{P}_1 = (-1)^{l_2} \eta'_{3} ,
\label{Z2-SU8}
\end{eqnarray}
where $\eta_{3}$ and $\eta'_{3}$ are the intrinsic $Z_2$ parities.
We assume that the $Z_2$ parity (\ref{Z2-SU8}) is assigned for the left-handed Weyl fermions.
The corresponding right-handed ones have opposite $Z_2$ parities.
Let us take $\eta_{3} = -1$ and $\eta'_{3} = -1$.
In this case, particles with even $Z_2$ parities are given in Table \ref{T5}.
Each particle possesses a zero mode whose scalar component is identified with one of the MSSM particles in 4 dimensions. 
In the second column, the quantum numbers after the charge conjugation are listed for the right-handed ones.
The subscript indicates the $U(1)_3$ charge.
In the last column, our particle identification is given for scalar partners.
Note that the particle identification is not unique but can be fixed by experiments.

\begin{table}[t]
\caption{Sfermions with even $Z_2$ parity from $[8, 3]$ with $(p, q, r, s)=(3, 2, 3, 0)$}
\label{T5}
\begin{tabular}{c|c|c} \hline
Rep. & Rep. for left-handed fermions & Sfermion species \\ \hline
$\left({}_{3}C_3, {}_{2}C_{0}, {}_{3}C_0\right)_R$ & $({\bf 1}, {\bf 1}, {\bf 1})_{-9}$ & $\tilde{e}^*_R$ \\
$\left({}_{3}C_1, {}_{2}C_{2}, {}_{3}C_0\right)_R$ & $(\overline{\bf{3}}, {\bf{1}}, {\bf{1}})_{-9}$ & $\tilde{u}^*_R$ \\
$\left({}_{3}C_1, {}_{2}C_{1}, {}_{3}C_1\right)_L$ & $({\bf 3}, {\bf 2}, {\bf 3})_{1}$ 
& $\tilde{q}_{1L}$, $\tilde{q}_{2L}$, $\tilde{q}_{3L}$ \\
$\left({}_{3}C_1, {}_{2}C_{0}, {}_{3}C_2\right)_R$ & $(\overline{\bf{3}}, {\bf{1}}, {\bf{3}})_{7}$ 
& $\tilde{b}^*_R$, $\tilde{s}^*_R$, $\tilde{d}^*_R$ \\ \hline
\end{tabular}
\end{table}

After the breakdown of $SU(3) \times U(1)_3$ gauge symmetry, we have the following mass formulae at $M_U$:
\begin{eqnarray}
&~& {m}_{\tilde{e}^*_R}^{2}(M_U) = {m}_{\tilde{u}^*_R}^{2}(M_U) = m_{[8,3]}^{2} - 9 D' ,
\label{SU(8)-e}\\
&~& {m}_{\tilde{q}_{1L}}^{2}(M_U) = m_{[8,3]}^{2} + D_1 + D_2 + D' ,
\label{SU(8)-q1}\\
&~& {m}_{\tilde{q}_{2L}}^{2}(M_U) = m_{[8,3]}^{2} - D_1 + D_2 + D' ,
\label{SU(8)-q2}\\
&~& {m}_{\tilde{q}_{3L}}^{2}(M_U) = m_{[8,3]}^{2}  - 2 D_2 + D' ,
\label{SU(8)-q3}\\
&~& {m}_{\tilde{b}^*_{R}}^{2}(M_U) = m_{[8,3]}^{2} + D_1 + D_2 +7 D' ,
\label{SU(8)-b}\\
&~& {m}_{\tilde{s}^*_{R}}^{2}(M_U) = m_{[8,3]}^{2} - D_1 + D_2 +7 D' ,
\label{SU(8)-s}\\
&~& {m}_{\tilde{d}^*_{R}}^{2}(M_U) = m_{[8,3]}^{2}  - 2 D_2 +7 D' ,
\label{SU(8)-d}
\end{eqnarray}
where $m_{[8,3]}$ is a soft SUSY breaking scalar mass parameter, $D_1$ and $D_2$ are parameters 
which represent $D$-term condensations related to the $SU(3)$ generator
and $D'$ stands for the $D$-term contribution of $U(1)_3$.
By eliminating these four unknown parameters, we obtain the relations
\begin{eqnarray}
&~& {m}_{\tilde{e}^*_R}^{2}(M_U) = {m}_{\tilde{u}^*_R}^{2}(M_U) ,~
\label{SU(8)-1} \\
&~& {m}_{\tilde{q}_{1L}}^{2}(M_U) - {m}_{\tilde{b}^*_{R}}^{2}(M_U) 
= {m}_{\tilde{q}_{2L}}^{2}(M_U) - {m}_{\tilde{s}^*_{R}}^{2}(M_U) = {m}_{\tilde{q}_{3L}}^{2}(M_U) 
- {m}_{\tilde{d}^*_{R}}^{2}(M_U) ,~ 
\label{SU(8)-2} \\
&~& 9 {m}_{\tilde{u}^*_{R}}^{2}(M_U) + 5 \left({m}_{\tilde{b}^*_{R}}^{2}(M_U) 
 + {m}_{\tilde{s}^*_{R}}^{2}(M_U) + {m}_{\tilde{d}^*_{R}}^{2}(M_U)\right) 
= 8 \left({m}_{\tilde{q}_{1L}}^{2}(M_U) + {m}_{\tilde{q}_{2L}}^{2}(M_U) 
+ {m}_{\tilde{q}_{3L}}^{2}(M_U)\right) .
\label{SU(8)-3}
\end{eqnarray}
Then using ordinary RG equations in the MSSM, we obtain the following sum rules among sfermion masses:
\begin{eqnarray}
&~& M_{\tilde{u}_{R}}^{2} - M_{\tilde{e}_{R}}^{2} 
= \zeta_3 M_3^2-20\zeta_1 M_1^2 
+\left(-\frac{5}{3}M_{W}^{2}+\frac{5}{3}M_{Z}^{2}\right)\cos 2\beta -10 \mathcal{S} ,
\label{SU8-1SR}\\
&~& M_{\tilde{u}_{L}}^{2} - M_{\tilde{b}_{R}}^{2} - 2F_b =  M_{\tilde{c}_{L}}^{2} - M_{\tilde{s}_{R}}^{2} 
 = M_{\tilde{t}_{L}}^{2} - M_{\tilde{d}_{R}}^{2} + F_t + F_b - m_t^2 ,
\label{SU8-2SR}\\
&~& 9 M_{\tilde{u}_{R}}^{2} + 5\left(M_{\tilde{b}_{R}}^{2} + M_{\tilde{s}_{R}}^{2} + M_{\tilde{d}_{R}}^{2}\right) 
 - 8\left(M_{\tilde{u}_{L}}^{2} + M_{\tilde{c}_{L}}^{2} + M_{\tilde{t}_{L}}^{2}\right) 
\nonumber \\
&~&  ~~~ =  - 24 \zeta_2 M_2^2 + 180 \zeta_1 M_1^2 +\left(-17M_{W}^{2}+5M_{Z}^{2}\right)\cos 2\beta 
+ 8 F_t - 2 F_b - 8 m_t^2 -30 \mathcal{S} ,
\label{SU8-3SR}
\end{eqnarray}
where $M_{\tilde{f}}^2$ represents the diagonal elements of the sfermion mass-squared matrices at the TeV scale, 
$M_i$ $(i=1, 2, 3)$
are the gaugino masses at the TeV scale, $\beta$ is defined 
in terms of the ratio of the VEVs of neutral components of the Higgs bosons as
$\tan \beta \equiv v_2/v_1$, and $F_t$ and $F_b$ stand for the effects of the top and bottom Yukawa interactions, respectively.
The parameters $\zeta_i$ and $\mathcal{S}$ are defined by
\begin{eqnarray}
&~& \zeta_3 \equiv -\frac{8}{9} \left(\left(\frac{\alpha_3(M_U)}{\alpha_3}\right)^2 -1\right) , ~~
\zeta_2 \equiv \frac{3}{2} \left(\left(\frac{\alpha_2(M_U)}{\alpha_2}\right)^2 -1\right) , ~~
\zeta_1 \equiv \frac{1}{198} \left(\left(\frac{\alpha_1(M_U)}{\alpha_1}\right)^2 -1\right) ,
\label{zeta}\\
&~& \mathcal{S} \equiv \frac{1}{10 b_1} \left(1- \frac{\alpha_1(M_U)}{\alpha_1}\right) 
\sum_{\tilde{F}} Y(\tilde{F}) n_{\tilde{F}} m_{\tilde{F}}^2,
\label{S}
\end{eqnarray}
where the quantities $\alpha_i \equiv g_i^2/(4\pi)$ are the structure constants defined 
by the gauge couplings $g_i$ at the TeV scale, 
$Y(\tilde{F})$ and $n_{\tilde{F}}$ represent the hypercharge and the degrees of freedom
of the sfermions and Higgs bosons $\tilde{F}$.

In the case with $\eta_{3} = -1$ and $\eta'_{3} = +1$, the following sum rules are obtained 
\begin{eqnarray}
&~& M_{\tilde{u}_{R}}^{2} - M_{\tilde{e}_{R}}^{2} = M_{\tilde{c}_{R}}^{2} - M_{\tilde{\mu}_{R}}^{2} =
M_{\tilde{t}_{R}}^{2} - M_{\tilde{\tau}_{R}}^{2} - m_t^2 + 2F_t - 2F_{\tau} 
\nonumber \\
&~& ~~~~~~ = \zeta_3 M_3^2-20\zeta_1 M_1^2 
+\left(-\frac{5}{3}M_{W}^{2}+\frac{5}{3}M_{Z}^{2}\right)\cos 2\beta -10 \mathcal{S} ,
\label{SU8-1SR'}\\
&~& M_{\tilde{u}_{R}}^{2} - M_{\tilde{\tau}_{L}}^{2} - F_{\tau} =  M_{\tilde{c}_{R}}^{2} - M_{\tilde{\mu}_{L}}^{2} 
 = M_{\tilde{t}_{R}}^{2} - M_{\tilde{e}_{L}}^{2} + 2 F_t - m_t^2 ,
\label{SU8-2SR'}\\
&~& 9 M_{\tilde{u}_{L}}^{2} + 5\left(M_{\tilde{e}_{L}}^{2} + M_{\tilde{\mu}_{L}}^{2} 
+ M_{\tilde{\tau}_{L}}^{2}\right) 
 - 8\left(M_{\tilde{u}_{R}}^{2} + M_{\tilde{c}_{R}}^{2} + M_{\tilde{t}_{R}}^{2}\right) 
\nonumber \\
&~&  ~~~~~~ =  -15\zeta_3 M_3^2 + 24 \zeta_2 M_2^2 - 240 \zeta_1 M_1^2 +\left(7M_{W}^{2} -10 M_{Z}^{2}\right)\cos 2\beta 
- 5 F_{\tau} + 16 F_t - 8 m_t^2 + 60 \mathcal{S} .
\label{SU8-3SR'}
\end{eqnarray}
Here we have used the particle identification such that $\left({}_{3}C_2, {}_{2}C_{1}, {}_{3}C_{0}\right)^c_R = \tilde{q}_{1L}$,
$\left({}_{3}C_2, {}_{2}C_{0}, {}_{3}C_{1}\right)_L$ $= \tilde{u}^*_{R}, \tilde{c}^*_{R}, \tilde{t}^*_{R}$,
$\left({}_{3}C_0, {}_{2}C_{2}, {}_{3}C_{1}\right)_L$ $= \tilde{e}^*_{R}, \tilde{\mu}^*_{R}, \tilde{\tau}^*_{R}$
and $\left({}_{3}C_0, {}_{2}C_{1}, {}_{3}C_{2}\right)^c_R$ $= \tilde{l}_{3L}, \tilde{l}_{2L}, \tilde{l}_{1L}$.
Here, the superscript $(c)$ represents the complex conjugate.

In the same way, we can derive relations among sfermion masses
on the basis of orbifold family unification models\cite{K&Kinami}. We find that the sum rules 
can be powerful probes of orbifold family unification, 
because they depend on the $Z_2$ parity assignment and the particle identification.

\section{Conclusions and discussion}

Orbifold SUSY GUTs possess excellent features such that the reduction of gauge symmetry is realized 
without fine-tuning among parameters related to Higgs masses 
and the proton stability is guaranteed by $U(1)_R$ symmetry.
Hence they are hopeful as a realistic model for the grand unification.
But we have the problem called the $\lq$arbitrariness problem', i.e.,
$\lq\lq$What is an origin of non-trivial $Z_2$ parities?".
Using the Hosotani mechanism, we find that theories are equivalent if the BCs are connected to by
gauge transformations and are classified into the equivalence classes of BCs.
Then the problem is restated as $\lq\lq$what is the principle to select a realistic equivalence class?".
One possibility is a dynamical determination of BCs in the framework of a fundamental theory.
I hope that an underlying theory must answer the question.

We also have tackled the origin of three families using $Z_2$ orbifolding 
and found that there are many models with three families of $SU(5)$ multiplets 
derived from a unique representation of $SU(N)$, but
no model with three families of the SM multiplets.
The riddle of family replication can be also solved by a fundamental theory.

Much work would be required to solve problems 
and to arrive at our goal: the construction of a realistic model.

\section*{Acknowledgements}
This work was supported in part by Scientific Grants from the Ministry of Education and Science, 
Nos.~18204024 and 18540259.

\bibliographystyle{plain}

\end{document}